\documentstyle[preprint,amsfonts,amssymb,aps,12pt]{revtex}
\input{epsf}
\tighten
\begin{document}
\draft
%
%  Titel and authors
%
\title{Reduction of the Glass Transition Temperature in Polymer Films:
A Molecular-Dynamics Study}
\author{F. Varnik$^1$%
\footnote{To whom correspondence should be addressed. Email:
{\sf varnik@mail.uni-mainz.de}},
J. Baschnagel$^2$, K. Binder$^1$\\[2mm]}
\address{$^1$Institut f\"ur Physik, Johannes-Gutenberg Universit\"at, D-55099 Mainz, Germany}
\address{$^2$Institut Charles Sadron, 6 rue Boussingault, F-67083, Strasbourg Cedex, France}
\maketitle
%
%       Commands for LaTex2e
%
%\newcommand{\mr}[1]{{\mathrm{#1}}}
%
%       Commands for LaTex 2.09
%
\newcommand{\mbf}[1]{{\mbox{\boldmath$#1$\unboldmath}}}                
\newcommand{\myfrac}[2]{{\frac{ \displaystyle {#1}} {\displaystyle {#2}}}}
\newcommand{\mr}[1]{{\rm #1}}
\newcommand{\mrm}[1]{{\rm #1}}
\newcommand{\myeq}{\!=\!}
\newcommand{\myapprox}{\!\approx\!}
\newcommand{\kB}{k_{\rm B}}
\newcommand{\Tc}{T_{\rm c}}
\newcommand{\Tnull}{T_{\rm 0}}
\newcommand{\Tcbulk}{T^{\rm bulk}_{\rm c}}
\newcommand{\Tnullbulk}{T^{\rm bulk}_{\rm 0}}
\newcommand{\Tg}{T_{\rm g}}
\newcommand{\PNext}{P_{\rm N,ext}}
\newcommand{\PText}{P_{\rm T,ext}}
\renewcommand{\vec}[1]{\mbox{\boldmath$#1$\unboldmath}}
%
%\receipt{ }
%
% ------------------------------------------------
% Abstract
% ------------------------------------------------
\begin{abstract}
We present results of molecular dynamics (MD)
simulations for a non-entangled polymer melt
confined between two completely smooth and repulsive walls, interacting 
with inner particles via the potential
$U_{\rm wall}\myeq (\sigma/z)^9$, where 
$z \myeq |z_{\rm particle}-z_{\rm wall}|$ 
and $\sigma$ is (roughly) the monomer diameter.
The influence of this confinement on the dynamic 
behavior of the melt is studied for various film thicknesses (wall-to-wall
separations) $D$, ranging from about 3 to about 14 times the bulk radius of gyration.
A comparison of the mean-square displacements in the film and in the bulk
shows an acceleration of the dynamics due to the presence
of the walls. 
This leads to a reduction of the critical temperature, $T_{\rm{c}}$,
of mode-coupling theory with decreasing film thickness.
Analyzing the same data by the Vogel-Fulcher-Tammann equation,
we also estimate the VFT-temperature $T_{\rm{0}}(D)$.
The ratio $T_{\mr{0}}(D)/T^{\mr{bulk}}_{\mr{0}}$ decreases 
for smaller $D$ similarly to $T_{\mr{c}}(D)/T^{\mr{bulk}}_{\mr{c}}$.
These results are in qualitative agreement with that of the
glass transition temperature observed in some experiments on supported 
polymer films.
\end{abstract}
%
%  Include PACS numbers, and Journal, to which paper will be submitted.
%
\pacs{{\sf PACS}: 61.20.Ja,61.25.Hq,64.70.Pf\\%
submitted to {\em Phys. Rev. E}
}
% ------------------------------------------------
% Text of the paper
% ------------------------------------------------
%%%%%%%%%%%%%%%%%%%%%%%%%%%
%%%%%%%%%%%%%%%%%%%%%%%%%%%
\section{Introduction}
\label{section::introduction}
%%%%%%%%%%%%%%%%%%%%%%%%%%%
%%%%%%%%%%%%%%%%%%%%%%%%%%%
Polymer science has had a major impact on the way we live.
Just 50 years ago, materials we now take for granted were 
non-existent. Due to their structural complexity, polymers 
are generally not crystalline at low temperatures. 
Rather they exhibit an amorphous, glassy structure.
The concept of the glass transition thus plays an important role in 
understanding the properties of polymer systems. Polymers are 
often used as protective coatings in 
microelectronics~\cite{Zhang::PolEngSci::1999,%
Rayss::JApplPolSci::1993,%
Armstrong::ElectrochemicaActa::1993}.
In such applications, the polymer is confined in a film geometry.
An important information for materials design is therefore 
how thermal properties of a polymer system are affected by 
the film geometry, in particular, whether and how the 
glass transition temperature,  $\Tg$, is influenced 
by confinement. 

In addition to its technological importance, the 
investigation of the  glass transition in thin polymer 
films is also from a great theoretical interest.
This is closely related to the unresolved nature of the
glass transition. Despite considerable experimental and
theoretical efforts, there is still not a fully 
satisfactory description of this phenomenon.
Phenomenological theories of the glass transition, 
such as the free volume theory~\cite{Cohen-Grest::PhysRevE1979,%
Cohen-Turnbull::JCP::31,%
Cohen-Turnbull::JCP::34,%
Cohen-Turnbull::JCP::52},
are attractive as they give a simple description of the 
basic observations but they contain adjustable parameters
whose physical significance is unclear.
Thermodynamic approachs, like 
Gibbs-DiMarzio theory~\cite{Gibbs-DiMarzio::JCP28::373,%
Gibbs-DiMarzio::JCP28::807}, treat the glass transition
as an ordinary thermodynamic phase transition but are faced with
the problem that the extrapolation of the theoretical results
leads to unphysical predictions like the existence of a temperature range
for which the system entropy becomes negative (Kauzmann paradoxon).

The so-called mode-coupling theory (MCT) is perhaps 
the most sucessfull of all descriptions of the glass transition.
Within this theory, there is a critical temperature, $\Tc$, 
at which the (structural) relaxation times diverge while the static
properties of the system remain liquid-like. The system  
freezes at $\Tc$ keeping its amorphous structure.
Thus, from the point of view of the MCT, the glass transition
is a purely dynamic phenomenon~\cite{Bengtzelius::JPhysC17,%
Goetze::LesHouches::1989,%
Goetze:Liquids::1991,%
Goetze-Sjoegren::TransportTheoryStatPhys,%
Goetze-Sjoegren::RepProgPhys55}.

However, it must be stressed that neither MCT nor other
descriptions of the glass transition have a definitive character.
One is therefore interested in finding ways to verify the 
basic concepts of different approachs.

Let us assume that the slowing down of the dynamics
when approaching $\Tg$ can be interpreted in terms of 
the critical slowing down of the dynamics near a 
second order phase 
transition~\cite{Kob::AnnualReviewCompPhys::Singapore95}.
A second order phase transition is usually characterized 
by the divergence of a length scale. The idea of a diverging 
length scale related to the glass transition has given 
rise to the concept of ``cooperative motion'' empirically 
introduced by Adam and Gibbs~\cite{Adam-Gibbs::JCP::43}.
According to Adam and Gibbs, near the glass transition,
individual particle motion is frozen out. Thus, the only 
possibility for structural relaxation is that of 
the collective motion of many particles. Note that,
here, the focus is no longer on the static correlations,
but on correlations between the \emph{dynamics} of particles.
The associated length scale is thus a dynamic one.
Let $\xi(T)$ denote the typical size of a cluster of
cooperatively moving particles. It is well-known that
close to a second order phase transition the relaxation time
of such a cluster scales like $\xi^z$, where $z\!>\!0$
is the so called dynamic critical 
exponent~\cite{Chaikin-Lubensky::PrinciplesofCondesedMatterPhysics}.
The sharp rise of the relaxation times near 
$\Tg$ is then explained by assuming the divergence of 
$\xi$ when lowering the temperature towards $\Tg$.

The above reasoning, however, is based on an empirical
assumption that $\xi(T)$ increases with decreasing temperature.
A significant improvement was achieved by Edwards and 
Vilgis~\cite{Edwards-Vilgis::PhysicaScriptaT13}.
These authors introduced an exactly solvable 
model system exhibiting glassy behavior at low $T$
and showed that the concept of cooperative motion 
alone was enough to give raise to a 
Vogel-Fulcher-Tammann (VFT)-law.

Recent computer experiments also 
support the idea of cooperative motion.
For example, it was observed by 
Kob \emph{et al.\ }~\cite{Kob-Donati-Plimpton-Poole-Glotzer::PRLett79,%
Donati-Glotzer-Poole-Kob-Plimpton::PRE60}
that particles move mainly in string-like clusters.
Bennemann \emph{et 
al.\ }~\cite{Bennemann-Donati-Baschnagel-Glotzer::Nature399}
report on a growing length scale
for the dynamics of a polymer
melt. Strong heterogeneity in the dynamics has also been 
observed in molecular dynamics simulations of 
bond breakage processes~\cite{Yamamoto::PRE::58::1998}
and of binary mixture of soft 
spheres~\cite{Yamamoto::JPhysSocJpn66,%
Yamamoto::EurophysLett40,%
Onuki-Yamamoto::JNoNCrystSolids235-237}.
It is found that particles move preferably 
within mobile clusters thus leading to 
a heterogeneity in the dynamics.

While easily detectable in a simulation, the regions of
heterogeneous dynamics are unfortunately not as easily 
accessible  to real experiments.
This results from the fact that dynamic heterogeneity is not
necessarily strongly correlated to density fluctuations.
The structures of these regions are therefore
more or less identical. Thus, one cannot use scattering 
experiments to determine the length scale $\xi(T)$~\cite{Forrest::2000}.
Fortunately, there is an issue to this problem.
Recall that the relaxation time of a cluster of 
strongly correlated particles such as
that observed in cooperative motion
scales like $\xi^z$, where $z\!>\!0$.
As the temperature decreases $\xi(T)$ becomes larger
and eventually reaches the system size, 
$\xi(T) \myeq L$. If this occurs, the relaxation time, $\tau$, of 
the system will scale like $\tau \propto L^z$. 
In other words, the relaxation dynamics of the system 
will become size 
dependent. This size dependence is indeed observed 
in Monte Carlo studies of the so called bond-fluctuation model 
(BFM) in 2D~\cite{Ray-Binder::EurophysicsLetters::27::1994}.
In the mentioned reference, the system size was varied while 
maintaining the periodic 
boundary conditions (pbc). An acceleration of the 
dynamics of the smaller systems have been observed in 
accordance with $\tau \propto L^z$.

A simple way of changing the system size in a real 
experiment is, for example, to vary the thickness 
of a planar film. Applying the same arguments as given above
to a thin film of thickness $D$ we should expect
finite-size effects on the dynamics for temperatures 
at which $\xi(T)\approx D$. Note that  boundary conditions 
are no longer periodic, but can change from an absorbing 
one to an approximately neutral or a repulsive one.

In this context, experiments on (model) systems reveal a 
stratified phenomenology. If the
interaction between the polymers and the substrate is attractive, 
the glass transition temperature $T_\mr{g}$ of the films becomes 
larger than the bulk value for small film
thicknesses~\cite{Keddie::FaradayDiscuss98}.
Intuitively, this effect can be attributed to chains which are close 
enough to the substrate to `feel' the attractive interaction.
The motion of these chains should be slowed down with respect to the bulk. 
In a thin film almost all chains touch the attractive 
substrate. So, $T_\mr{g}$ should increase.

On the other hand, measurements (by ellipsometry) of $\Tg$ for
polystyrene (PS) films (of rather large molecular weights)
on a silicon substrate showed a significant decrease of $\Tg$ 
from $375{\rm K}$ down to $345{\rm K}$ for the smallest film 
thickness of $10$nm, i.e. a relative change of $10\%$ in $\Tg$ was 
observed~\cite{Keddie::Jones::Cory::EuroPhysLett28}. There have also been 
many experiments in recent years on freely standing 
polystyrene films (i.e., no solid substrate, but two polymer-air 
interfaces)~\cite{Forrest::PRL77::page4108,%
Forrest::PRL77::page2002,%
ForrestJones2000,%
Forrest::PRE56} showing a dramatic 
decrease of $T_\mr{g}$ by up to $20\%$ if the film thickness 
becomes much smaller than the chain size. An interesting explanation 
of this observation in terms of an interplay between polymer-specific properties and
free-volume concepts has been proposed~\cite{deGennses::EurPhysJE2000}.
This decrease becomes much weaker if one or even both of the free interfaces 
are replaced by a weakly interacting solid substrate. Whereas the strong depression 
of $T_\mr{g}$ in the freely standing film could possibly be attributed to the 
significant release of geometric constraints at the air-polymer interface, 
the acceleration of the structural relaxation of a polymer melt between two (almost) 
neutral solid substrate is much harder to understand intuitively.

Recent simulation results emphasize the fact that the increase or 
decrease of the glass transition temperature does strongly 
depend on the  interaction between the polymer 
chains and the substrate~\cite{Torres-Nealey-dePablo::PRL85}.
For a model of square-well spherical interaction 
sites interconnected by fully flexible strings one observes a 
reduction of $\Tg$ in the case of a weak attraction between the 
substrate and the chains, whereas for the case of strong 
attraction an increase of $\Tg$ is 
found~\cite{Torres-Nealey-dePablo::PRL85}.

We study a continuum model where polymer chains are confined
between two identical, ideally flat and purely repulsive 
walls~\cite{Varnik-Baschnagel-Binder::JPhysIV10::ConfProc,%
Varnik-Baschnagel-Binder::JCP113::2000}, obtaining complete information 
in both space and time in atomistic detail. Instead of focusing on
a computation of $\Tg$, we rather check some predictions of the 
mode-coupling theory  and investigate the dependence of
the mode-coupling critical temperature, $\Tc$, on film thickness $D$.
We also examine the $D$-dependence of the VFT-temperature, $T_0$, 
with the result that the ratios 
$T_0(D)/\Tnullbulk$ and $\Tc(D)/\Tcbulk$ behave similarly.

After a presentation of the model in the next section, we discuss 
in section~\ref{section::influence::of::ensembles::on::the::dynamics}
the reliability of the system dynamics obtained from molecular 
dynamics (MD) simulations within $NVT$ and $NpT$ ensembles. In 
section~\ref{section:Dependence_of_Tc_on_Film_Thickness}
we present MD results on the influence of the 
walls on the system mobility and its critical behavior at low 
temperatures for a variety of film thicknesses showing 
that the critical temperature is lowered for stronger confinements 
(smaller film thicknesses). Section~\ref{section:Local_Dynamics}
is devoted to a brief analysis of the local dynamics where we address a 
subtle point concerning the proper definition of local quantities
in terms of particle positions and/or momenta. 
A conclusion compiles the results.
%%%%%%%%%%%%%%%%%%%%%%%%%%%%%%%%%%%%%%%%%%%%%%%%%%%%%%%%%%%%%%%%%%%%
%%%%%%%%%%%%%%%%%%%%%%%%%%%%%%%%%%%%%%%%%%%%%%%%%%%%%%%%%%%%%%%%%%%%
%%%%%%%%%%%
%%%%%%%%%%%
\section{Model}
\label{section::model}
%%%%%%%%%%%
%%%%%%%%%%%
%%%%%%%%%%%%%%%%%%%%%%%%%%%%%%%%%%%%%%%%%%%%%%%%%%%%%%%%%%%%%%%%%%%%
We study a Lennard-Jones (LJ) model for a dense polymer
melt~\cite{Kremer-Grest::JCP92,Bennemann-Paul-Binder-Duenweg::PRE57}
of short chains (each consisting of 10 monomers) embedded between two completely
smooth, impenetrable 
walls~\cite{Varnik-Baschnagel-Binder::JPhysIV10::ConfProc,%
Varnik-Baschnagel-Binder::JCP113::2000}.
Two potentials are used for the interaction between particles. The first one is a 
truncated and shifted LJ-potential which acts between all pairs of particles
regardless of whether they are connected or not,
$$
U_{\mbox{\scriptsize LJ-ts}}(r)=\left\{ 
\begin{array}{ll}
U_{\mr{LJ}}(r)-U_{\mr{LJ}}(r_{\mr{c}}) & \mbox{if $r<r_{\mathrm c}$}\;, \\
0 & \mbox{otherwise}\;,
\end{array}
\right.
$$
where 
$U_{\mr{LJ}}(r)\myeq  4 \epsilon \Big[ (\sigma/r)^{12} -(\sigma/r)^{6}\Big]$
and $r_{\mr{c}}\myeq 2\times 2^{1/6}\sigma$.
The connectivity between adjacent monomers of a chain is ensured by a
FENE-potential~\cite{Kremer-Grest::JCP92},
\begin{equation}
U_{\mathrm{FENE}}(r)=-\frac{k}{2} R^2_0 \ln \bigg[1-\Big 
(\frac{r}{R_0}\Big)^2\bigg]
\;,
\label{eq:FENE::potential::def}
\end{equation}
where $k\myeq 30 \epsilon/\sigma^2$ is the strength factor 
and  $R_0\myeq 1.5\sigma$ the maximum 
allowed length of a bond. The wall potential was chosen as
\begin{equation}
U_{\mathrm W}(z)=\epsilon \bigg( \frac{\sigma}{z}\bigg )^9 \; ,
\label{eq:def:uw}
\end{equation}
where $z\myeq |z_{\mbox{\scriptsize particle}}-z_{\mbox{\scriptsize wall}}|$
($z_{\mbox{\scriptsize wall}}\myeq  \pm D/2$). This corresponds to 
an infinitely thick wall made of infinitely small particles which interact
with inner particles via the potential 
$45\epsilon(\sigma/r)^{12}/(4\pi\rho_{\mathrm{wall}}\sigma^3)$
where $\rho_{\mathrm{wall}}$ denotes the density of wall particles. The sum 
over the wall particles then yields $\epsilon (\sigma/z)^{9}$.
All simulation results are given in Lennard-Jones (LJ) units.
All lengths and energies are measured respectively in units of $\sigma$ and $\epsilon$,
temperature in units of $\epsilon/\kB$ ($\kB \!=\! 1$) and time in units 
of $(m\sigma^2/\epsilon)^{1/2}$.

\begin{figure}
\epsfxsize=105mm 
\hspace*{23mm}\epsffile{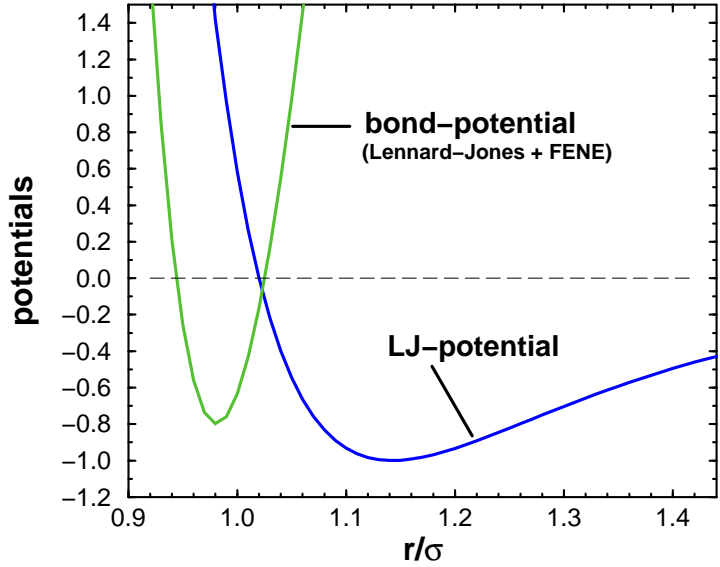}
\epsfxsize=100mm
\hspace*{25mm}\epsffile{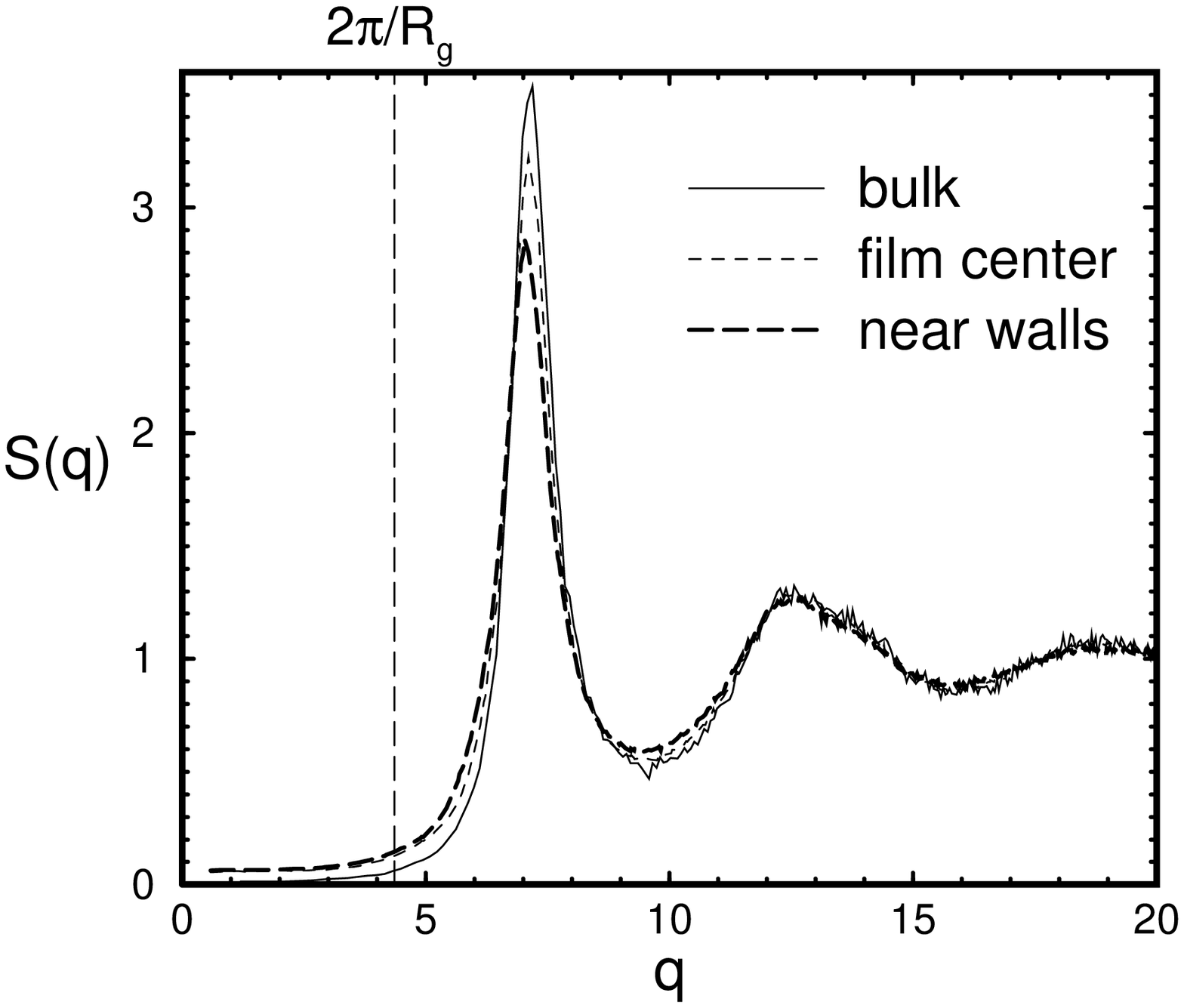}
\caption[]{Upper panel: Illustration of the potentials of the model.
The bond-potential results from a superposition of the Lennard-Jones 
(LJ) and the FENE potentials. The minimum position of the bond-potential 
is smaller than that of the LJ-potential. This incompatibility favors 
amorphous structure at low $T$ which is confirmed by the behavior of $S(q)$.
Lower panel: Comparison of the static structure factor $S(q)$ of the
melt in the bulk and in the film ($D\myeq 10$) at $T\myeq 0.46$ 
(critical temperature of mode-coupling
theory in the bulk:  $T_\mr{c} \simeq 0.45$ \cite{Bennemann-Baschnagel-Paul::EPJB10}). 
`Near walls' and `film center'
mean averages over the regions close to the walls ($0\leq z \leq 3.5$)
and the inner portion of the film ($3.5\leq z \leq 5$). $z$ is the 
distance of a particle from the (left) wall. The vertical
dashed line indicates the $q$-value corresponding to the 
bulk radius of gyration ($R_\mr{g}^2 \simeq 2.09$).}
\label{fig:model}
\end{figure}

The upper panel of Fig.~\ref{fig:model} compares the bond potential,
i.e.\ the sum of LJ- and FENE-potentials, with the LJ-potential.
It shows that the bonded monomers prefer 
shorter distances than the nonbonded ones. Thus, our model
contains two intrinsic length scales. Since these length scales
are chosen to be incompatible with a (fcc or bcc) crystalline structure,
one could expect that the system does not crystallize at
low temperatures, but remains amorphous. This expectation is well
borne out. The lower panel of 
Fig.~\ref{fig:model} the structure factor, $S(q)$, of the bulk system
with that of a film of thickness $D\myeq 10$ at $T\myeq 0.46$
(note that the mode-coupling critical 
temperature of the present model in the bulk is 
$T^{\rm bulk}_{\rm c}\myeq 0.45$~\cite{Bennemann-Baschnagel-Paul::EPJB10}).
For the film, two structure factors are chosen, corresponding to 
two different regions: a layer of thickness $3\sigma$ centered
in the middle of the film (called ``film center'') and the region
between this layer and the walls (called ``near walls''). In both cases,
$S(q)$ is calculated parallel to the walls [i.e.\ $q\myeq |q|,\;
\mbf{q}\myeq (q_x,q_y)$] by averaging over all monomers in 
the respective region.

Fig.~\ref{fig:model} shows that the structure of the melt in 
the bulk and in the film is characteristic of an amorphous material. 
At small $q$, $S(q)$ is small, reflecting the low compressibility 
of the system. Then, it increases and develops a peak at 
$q_{\rm max}$ which corresponds to the local packing of monomers 
($2\pi/q_{\rm max}\myapprox 1$) before it decreases again and 
begins oscillating around $1$, the large-$q$ limit of $S(q)$. 

The most prominant differences between the bulk and the film 
are found for small $q$ and for $q_{\rm max}$. The compressibility 
of the film is higher, the value of $q_{\rm max}$ is shifted to
slightly lower $q$ and the magnitude of $S(q_{\rm max})$ is smaller
than in the bulk. In the bulk one can observe a similar shift
of $q_{\rm max}$ and decrease of $S(q_{\rm max})$ as the temperature
increases. Also the compressibility of a bulk fluid increases 
with temperatures.  Therefore,  the  local  packing of the monomers
in the film seems to resemble that of the bulk at a 
higher temperature. Since the local structure of the melt has an important 
influence on its dynamic behavior in the supercooled 
state~\cite{Goetze::LesHouches::1989}, 
Fig.~\ref{fig:model} suggests that the film releases 
more easily than the bulk at the same temperature. Indeed, we will 
see later that the dynamics of the system is much faster in the 
film than in the bulk when compared at the same temperature.

All simulations have been carried out under constant normal
pressure $\PNext\myeq p \myeq  1$. However, to adjust the normal pressure, 
we do \emph{not} vary the wall-to-wall separation, $D$, but the 
surface area. For each temperature, the average surface area is 
calculated by an iterative 
approach~\cite{Varnik-Baschnagel-Binder::JCompPhys::2001}.
The system is then propagated untill the instantaneous 
surface area reaches
the computed average value. At this point the surface area (and thus the volume)
is fixed and a production run is started in $NVT$-ensemble, where
the system temperature is adjusted using the Nos{\'e}-Hoover 
thermostat~\cite{Nose::JCP81,Hoover::PhysRevA31}.
More details about the applied simulation techniques can be found 
in~\cite{Varnik-Baschnagel-Binder::JCP113::2000,%
Varnik-Baschnagel-Binder::JCompPhys::2001,%
Varnik::Dissertation::Mainz2000}.

%%%%%%%%%%%%%%%%%%%%%%%%%%%%%%%%%%%%%%%%%%%%%%%%%%%%%%%%%%%%%%%%%%%%
%%%%%%%%%%%
%%%%%%%%%%%
\section{$NVT$ and $NpT$ versus $NVE$: Influence on the Dynamics}
\label{section::influence::of::ensembles::on::the::dynamics}
%%%%%%%%%%%
%%%%%%%%%%%
%%%%%%%%%%%%%%%%%%%%%%%%%%%%%%%%%%%%%%%%%%%%%%%%%%%%%%%%%%%%%%%%%%%%
It was mentioned in section~\ref{section::model} that the production runs
were performed at constant volume and temperature using Nos\'e-Hoover thermostat.
This thermostat slows down or accelerates all particles
depending on the sign of the difference between the instantaneous
kinetic energy of the system and the desired value given by the imposed 
temperature, i.e.  $3N\kB T/2$ ($N$ is the number of 
particles)~\cite{Nose::JCP81,%
Hoover::PhysRevA31,%
Nose-Klein::MolPhys50,%
Hoover::PhysRevLett48,%
Hoover::PhysRevA34,%
Nose::MolPhys52}.
One may therefore ask how reliable the resulting dynamics is when 
compared to pure Newtonian dynamics in the microcanonical ($NVE$) ensemble.
This question was already examined for the bulk system 
in~\cite{Bennemann-Paul-Binder-Duenweg::PRE57}. In this case, the 
results obtained within constant energy ($NVE$) simulation and that 
using the Nos\'e-Hoover thermostat are identical.

We are going to show that the presence of the walls does not
change this behavior. However, we go a step further
and also investigate the influence of volume fluctuations on the system
dynamics. This point is very important if one is interested in constant
pressure simulations. We will see that, contrary to the case of 
Nos\'e-Hoover thermostat, the system dynamics is strongly perturbed
when the system volume is allowed to fluctuate
(therefore, for a given normal pressure, the corresponding average volume
was first computed for each $T$ in a $NpT$-simulation and then the 
dynamics was  analyzed in  production runs at constant volume).
For this purpose, we compare results obtained
from MD simulations within the $NVE$ (microcanonical), 
$NVT$ (canonical, using Nos\'e-Hoover thermostat) and 
$NpT$ (Nos\'e-Hoover thermostat plus the fluctuations of the 
surface area) ensembles. 

Note that all results to be discussed in this section 
correspond to a film of thickness $D\myeq 5$. Recall that $D$  
stands for the wall-to-wall separation. The distance of the 
closest approach of a monomer to a wall is approximately
its own diameter (i.e. $\sigma \myeq 1 $).
Therefore, a value of $D\myeq 5$ corresponds to the
extreme case of three monomer layers only.

Let us first consider the velocity autocorrelation function 
defined as
\begin{equation}
C_{v}(t)={\Big< \sum^{N}_{i} \mbf{v}_{i}(0) \mbf{v}_{i}(t)  \Big>}/
{\Big< \sum^{N}_{i=1} v^2_{i}(0) \Big>} \; .
\label{eq:VACF}
\end{equation}

\begin{figure}
\epsfxsize=140mm 
\hspace*{0mm}
\epsffile{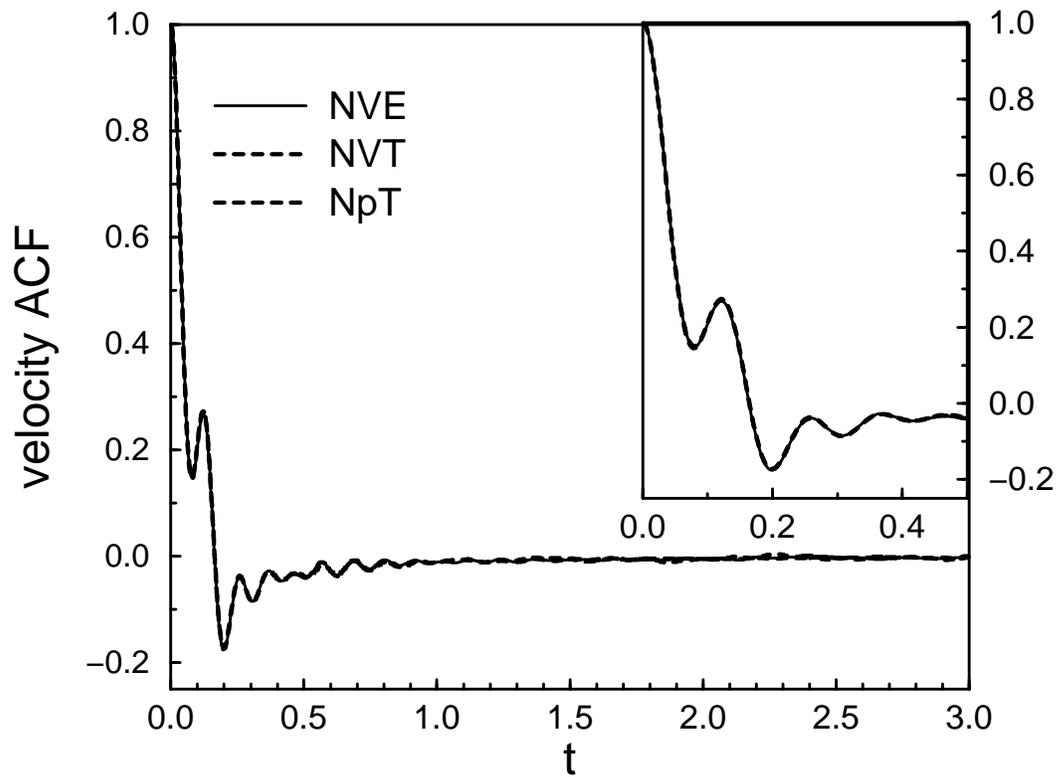}
\caption{
Velocity autocorrelation function (ACF) as computed from simulation runs
within the $NVE$, $NVT$ and $NpT$-ensembles. Neither the coupling to the heat bath nor
the fluctuations of the volume seem to affect the behavior of this quantity.
The inset shows a magnification of the initial behavior of the velocity-ACF
($D\myeq5,\; P_N\myeq 1,\; T\myeq 0.55,\; N\myeq 500$).}
\label{fig::VACF_T0.55_p1_D5_NVE_vs_NVT_and_NpT}
\end{figure}

\begin{figure}
\epsfxsize=140mm 
\hspace*{0mm}\epsffile{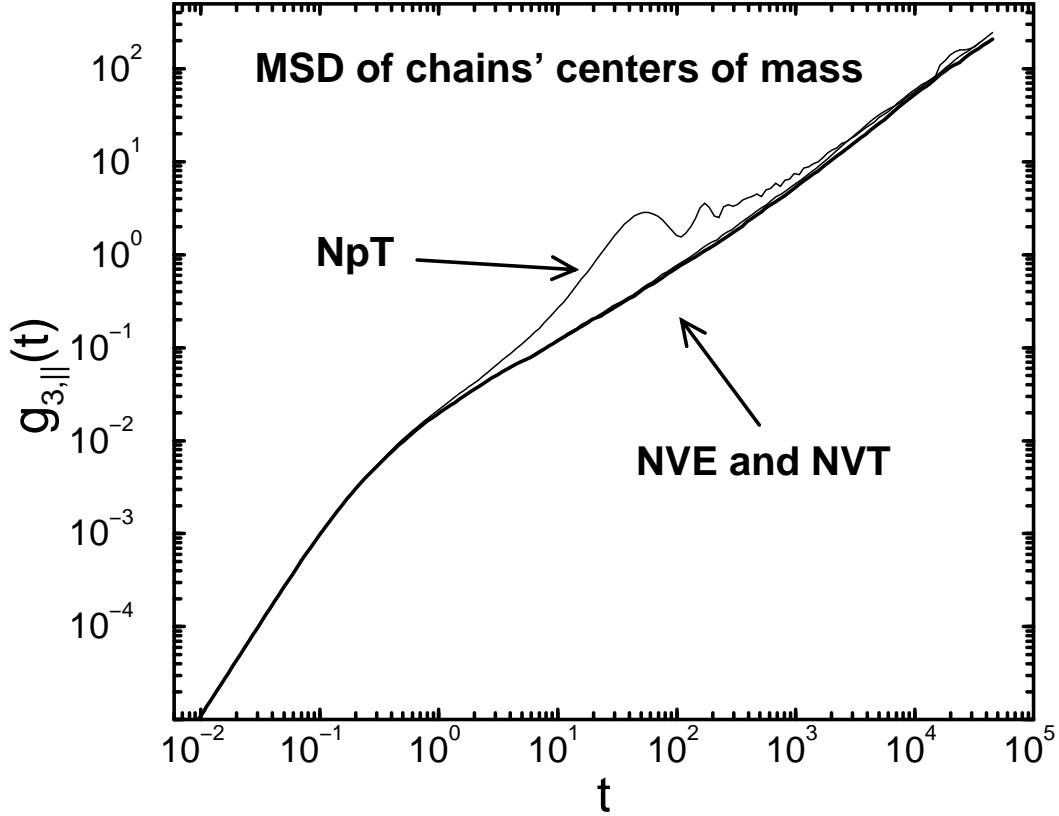}
\caption{
The mean-square displacement (MSD) of chain's center of mass 
in direction parallel to the walls,
$g_{3,\parallel}(t)$, obtained from $NVE$, $NVT$ and $NpT$ simulation runs
($D\myeq5,\; P_N\myeq 1,\; T\myeq 0.55,\; N\myeq 500$).
Obviously, the $NVT$-result is  identical to that obtained within the
$NVE$-ensemble simulation. Contrary to that, the  time evolution of
$g_{3,\parallel}(t)$ in the $NpT$-ensemble simulation is 
unphysical for $t\ge 1$.
}
\label{fig::g3_T0.55_p1_D5_NVE_vs_NVT_and_NpT}
\end{figure}

Figure~\ref{fig::VACF_T0.55_p1_D5_NVE_vs_NVT_and_NpT} presents
results for $C_{v}(t)$ obtained from simulations in the $NVE$, 
$NVT$, and $NpT$ ensembles. As seen from this figure, no difference 
is observed for  $C_{v}(t)$ within various ensembles.
The velocity autocorrelation function vanishes so 
rapidly that it can be equally well computed within all 
these ensembles. However, quantities that evolve slowly
in time exhibit a different behavior. An example is
the mean-square displacement (MSD) of chain's center of mass,
\begin{eqnarray}
g_{3, \parallel} (t) &=& \frac{1}{M} \sum^{M}_{i=1}
\left \langle
\big[ \mbf{R}_{\mrm{cm},i,\parallel}(t)-\mbf{R}_{\mrm{cm},i,\parallel} \big]^{2}
\right \rangle \; .
\label{eq:g3a::def}
\end{eqnarray}
Here,  $M$ is the number of chains 
and $\mbf{R}_{\mrm{cm},i,\parallel}$ is 
the projection of the center of mass vector of the $i$-th chain 
onto a plane parallel to the wall. 
Figure~\ref{fig::g3_T0.55_p1_D5_NVE_vs_NVT_and_NpT} shows that
data for $g_{3,\parallel}(t)$ within $NVE$ and $NVT$ ensembles
are identical.
However, the result obtained from the $NpT$-ensemble differs strongly from the 
reference $NVE$-curve. This discrepancy is due to the relative small
box size of a typical MD simulation. Note that the relative volume 
fluctuations scale as $1/\sqrt{N}$.  While fully negligible for real 
systems ($N\approx \! 10^{23}$), these fluctuations
become important in a simulation where the particle number 
is of order of~$1000$.

%%%%%%%%%%%%%%%%%%%%%%%%%%%%%%%%%%%%%%%%%%%%%%%%%%%%%%%%%%%%%%%%%%%%
%%%%%%%%%%%
%%%%%%%%%%%
\section{Dependence of $\Tc$ on Film Thickness}
\label{section:Dependence_of_Tc_on_Film_Thickness}
\begin{figure}
\hspace*{-15mm}
\epsfxsize=140mm
\hspace*{15mm}\epsffile{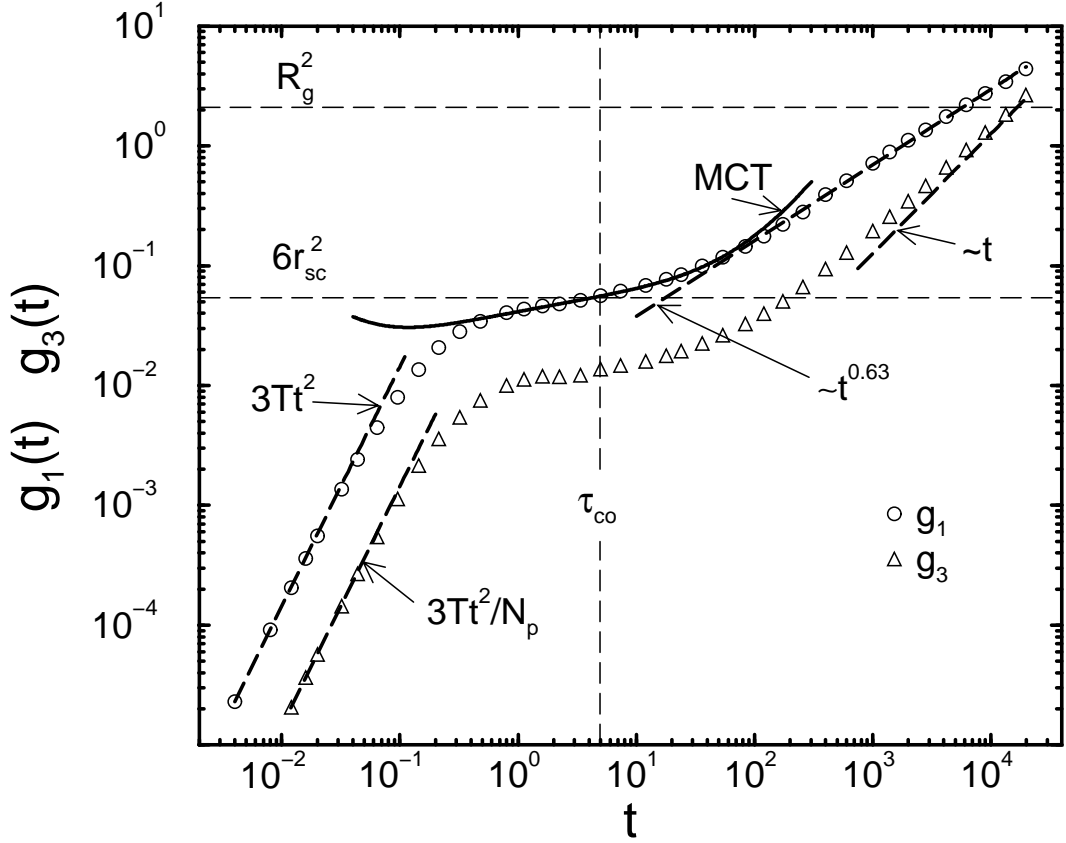}
\caption[]{Log-log plot of the mean-square displacements of the innermost 
monomer, $g_1(t)$,
and of the chain's center of mass, $g_3(t)$, versus time for $T\myeq 0.48$. 
The initial ballistic behaviors 
for $g_1(t)$ and $g_3(t)$, i.e., $g_1(t)\myeq \langle v^2\rangle t^2\myeq 3Tt^2$ and
$g_3(t)\myeq 3Tt^2/N_{\rm p}$ ($\langle v^2\rangle$: mean-square monomer 
velocity, $N_{\rm p}\myeq 10$: chain length),
and the late time diffusive behavior are indicated as dashed lines. In addition, 
a power law fit $g_1(t) \sim t^x$ with an effective exponent $x \simeq 0.63$ 
is shown as another dashed line. The dashed horizontal lines represent the 
radius of gyration $R_{\rm g}^2$ 
($\simeq 2.09$; upper line) and the plateau value $6r_{\rm sc}^2$ of a 
MCT-analysis ($\simeq 
0.054$; lower line), respectively. The dashed vertical line indicates the 
time, $\tau_\mr{co}$,
of the intersection between $g_1(t)$ and $6r_{\rm sc}^2$. Additionally, 
the mode-coupling 
approximation in the $\beta$-relaxation regime is shown as a thick solid 
line. Adapted from 
reference~\cite{Bennemann-Baschnagel-Paul-Binder}.}
\label{fig:g1.T0.48.bulk}
\end{figure}

%%%%%%%%%%%
%%%%%%%%%%%
%%%%%%%%%%%%%%%%%%%%%%%%%%%%%%%%%%%%%%%%%%%%%%%%%%%%%%%%%%%%%%%%%%%%
We now focus on the influence of the walls on the sluggish dynamics 
of the system. For this purpose, it is instructive to recall 
some important features of the present model in
the bulk at low temperatures. In Fig.~\ref{fig:g1.T0.48.bulk} 
the mean-square displacements of the innermost monomer,
\begin{eqnarray}
g_{1} (t) &=& \frac{1}{M} \sum^{M}_{i=1} \left \langle
\big[ \mbf{r}_{i, \mrm{inner}} (t)
- \mbf{r}_{i,\mrm{inner}}(0) \big]^{2}    \right \rangle \; ,
\label{eq:g1a::def}
\end{eqnarray}
and of the chain's center of mass, $g_3(t)$, are displayed versus 
time for $T\myeq 0.48$.
For short times, the motion of the system
can be described by assuming free particles (ballistic regime):
$g_1(t)\myeq \langle v^2\rangle t^2\myeq 3Tt^2$ and $g_3(t)\myeq 3Tt^2/N_{\rm p}$,
where $\langle v^2\rangle$ is the mean-square monomer velocity 
and $N_{\rm p}\myeq 10$ is the number of the monomers of a chain.
In agreement with the predictions of the mode-coupling 
theory (MCT)~\cite{Bengtzelius::JPhysC17,%
Goetze::LesHouches::1989,%
Goetze:Liquids::1991,%
Goetze-Sjoegren::TransportTheoryStatPhys,%
Goetze-Sjoegren::RepProgPhys55},
a plateau regime emerges after the ballistic motion.
The tagged particle ``feels'' the presence of its neighbors
and, as the temperature is low, remains temporarily in the cage
formed by these neighbors.
However, contrary to a simple (atomic) liquids, where a 
direct crossover from the plateau into the diffusive regime occurs,
an intermediate subdiffusive regime emerges due to the connectivity 
of the monomers~\cite{Rouse::JChemPhys21:1953}.
In this regime, the center of mass already crosses over to the
asymptotic diffusive motion, $g_3(t)\simeq t$, whereas the 
motion of the innermost monomer is described by a power 
law $g_1(t) \sim t^x$ with an effective exponent 
$x \simeq 0.63$. The innermost  monomers reach the
diffusive limit only if $g_1(t)$
is larger than the end-to-end distance of a chain and thus 
outside of the time window shown in the figure.

\begin{figure}
\epsfxsize=100mm
\hspace*{15mm}\epsffile{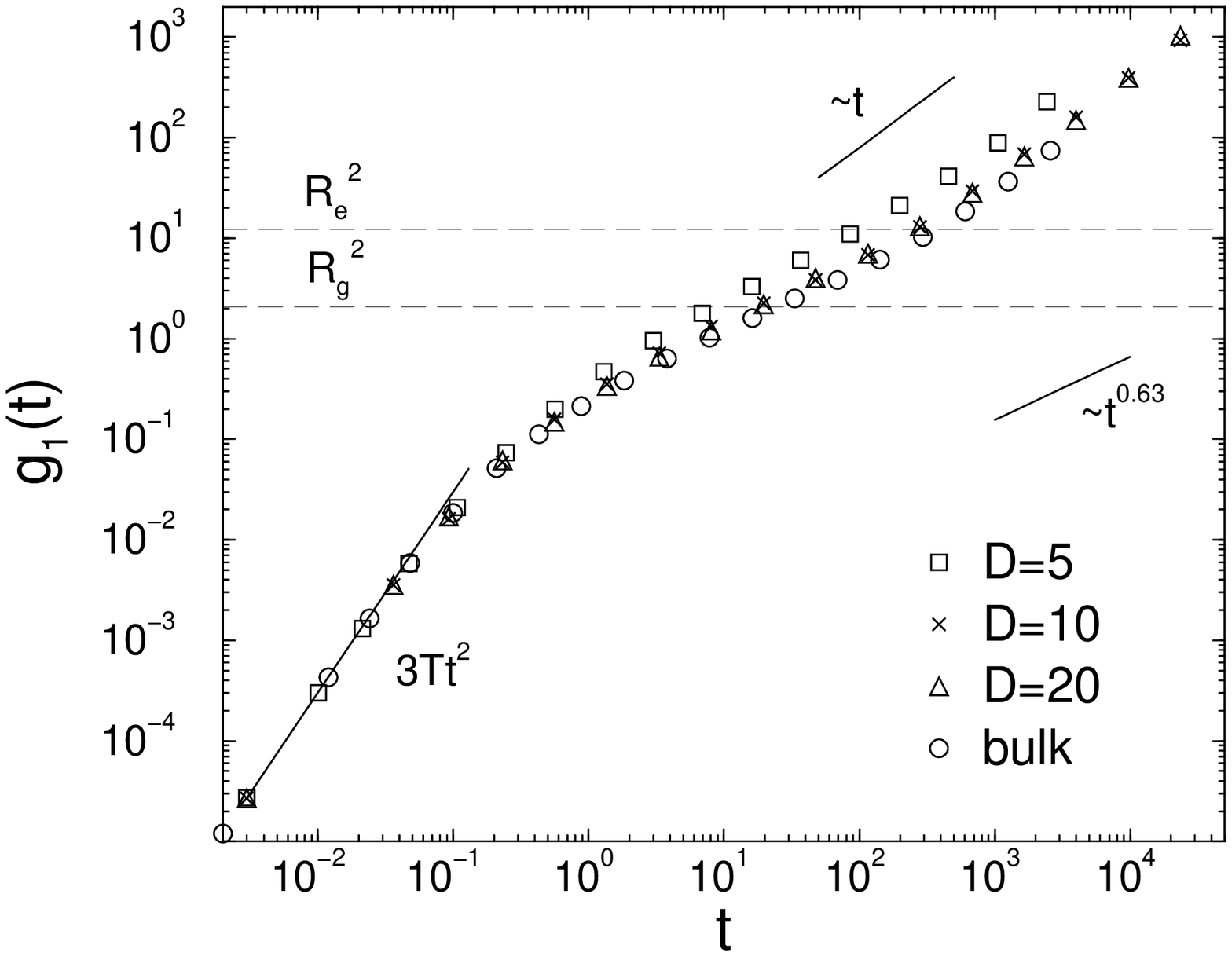}
\epsfxsize=102mm
%\epsfysize=102mm
%\hspace*{20mm}\epsffile{g1_compare_film_and_bulk.eps}
\hspace*{20mm}\epsffile{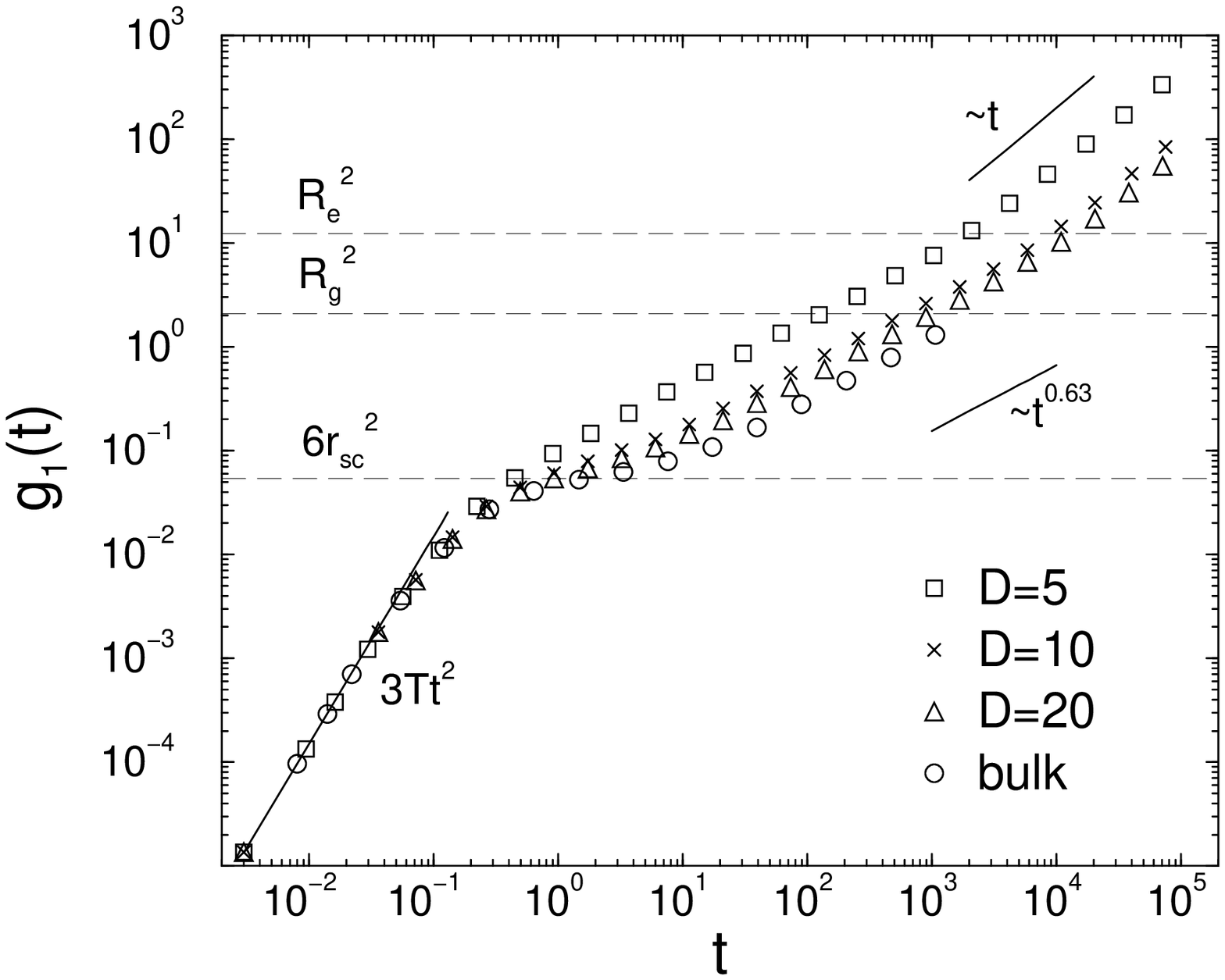}
\caption[]{Log-log plot of the mean-square displacement of the innermost 
monomer $g_1(t)$ for two different temperatures: $T\myeq 1$ (high 
temperature, normal liquid state; upper panel) and $T\myeq 0.5$ 
(supercooled state close to $T_\mr{c}^\mr{bulk}\myeq 0.45$; lower panel). 
The figures compare bulk
data with the displacements measured parallel to the walls in films 
of different thickness: $D\myeq 5$ ($\approx \! 3.5 R_\mr{g}$), 
$D\myeq 10$ ($\approx \! 7 R_\mr{g}$) and $D\myeq 20$ ($\approx \! 14 R_\mr{g}$). 
The film data were multiplied by $3/2$ to account for the 
different number of spatial directions
used to calculate $g_1(t)$ (i.e., 3 directions for the bulk, but only 2 for the
films). The bulk end-to-end distance $R^2_\mr{e}$ 
and radius of gyration $R_\mr{g}^2$ are 
indicated as dashed horizontal lines.  Furthermore, the lower panel also 
shows the plateau value $6r_{\rm sc}^2$ of a MCT-analysis ($\simeq 0.054$; 
lowest dashed line). The solid lines represent the behavior of $g_1(t)$ 
expected in different time regimes: ballistic at short times ($g_1(t)\sim t^2$), 
diffusive at late times ($g_1(t)\sim t$), and dominated by chain connectivity
for times where $g_1(t)> 1\myeq $ monomer diameter ($g_1(t)\sim t^x$; 
$x \myeq  0.63 \myeq $ effective exponent).}
\label{fig:msd.T1+0.5.D=5+10+20+bulk}
\end{figure}

In Fig.~\ref{fig:msd.T1+0.5.D=5+10+20+bulk} we compare the 
mean-square displacement of the innermost monomer $g_1(t)$ 
for films of various thicknesses with that of the bulk. 
The upper panel corresponds to 
a high temperature of $T\myeq 1$ where the system properties are liquid-like.
At this temperature the influence of the walls is rather small so that
$g_1(t)$ of the bulk almost overlaps with that of the film for thicknesses 
$D \!\ge\! 10$. However, the  lower panel of
Fig.~\ref{fig:msd.T1+0.5.D=5+10+20+bulk} shows that
the effect of the walls on the mobility becomes significant 
at all studied film thicknesses with progressive supercooling.
Outside the initial ballistic regime, the motion is faster, the
smaller the film thickness.

To quantify this observation, we define relaxation times
as the time needed by a given mean-square displacement
(like $g_3$, the MSD of chain's center of mass or $g_1$, the MSD of the 
innermost monomer) to reach the monomer size
\begin{eqnarray}
g_{i} (t \myeq \tau) &:=& 1 \mbox{~(defining equation for $\tau$).}
\label{eq:tau_gi::def}
\end{eqnarray}
Note that, due to the film geometry, there are only two independent 
directions (each of them parallel to the walls) for diffusive motion 
compared to three in a homogeneous melt. For the film, we thus 
compute $g_{i}$ using these two directions and then multiply 
the result by a factor of $3/2$. This multiplication is 
necessary if the film data are to be compared with the corresponding 
bulk quantities.

Using Eq.~(\ref{eq:tau_gi::def}), we computed $\tau(g_{i} \myeq 1)$ as a 
function of temperature for  various film thicknesses, where,
in addition to $g_3$ and $g_1$, the MSD of all monomers,
\begin{equation}
g_{\,0} (t) = \frac{1}{N} \sum^{N}_{i=1}
\left \langle \big [ \mbf{r}_{i}(t) - \mbf{r}_{i}(0) \big ]^{2} \right \rangle \; ,
\label{eq:g0a::def}
\end{equation}
and that of the end monomers (chain ends),
\begin{eqnarray}
g_{4} (t) &=& \frac{1}{M} \sum^{M}_{i=1} \left \langle
\big[ \mbf{r}_{i, \mrm{end}} (t)
- \mbf{r}_{i,\mrm{end}}(0) \big]^{2}    \right \rangle \;,
\label{eq:g4a::def}
\end{eqnarray}
have also been used.

For each film thickness, this yields four different relaxation 
times as a function of temperature.
As the mode-coupling theory~\cite{Goetze::LesHouches::1989,%
Goetze:Liquids::1991,%
Goetze-Sjoegren::RepProgPhys55,%
Goetze-Sjoegren::TransportTheoryStatPhys,%
Bengtzelius::JPhysC17}
has been rather 
successful in describing the slow dynamics of the present 
model in bulk~\cite{Bennemann-Baschnagel-Paul::EPJB10,%
Aichele-Baschnagel::EurPhysJE::I,%
Aichele-Baschnagel::EurPhysJE::II}, we tried to fit $\tau(T)$ via a 
power law,
\begin{equation}
\tau(T) \propto |T-\Tc(D)|^{-\gamma(D)}\; .
\label{eq:MCT::power::law::for::tau}
\end{equation}
Such a power-law divergence of the $\alpha$-relaxation time
is predicted by MCT for the bulk~\cite{Goetze::LesHouches::1989,%
Goetze:Liquids::1991,%
Goetze-Sjoegren::RepProgPhys55}.

The fit is done as follows: First, all (three) parameters of
Eq.~(\ref{eq:MCT::power::law::for::tau}) were adjusted. The values 
of $\gamma$ obtained from $\tau(g_{0}\myeq 1)$, $\tau(g_{1}\myeq 1)$,
$\tau(g_{3}\myeq 1)$ and $\tau(g_{4}\myeq 1)$ agreed well within 
the error bars. Therefore, we fixed $\gamma$ at the average 
value for the given film thickness and repeated the fits.

\begin{figure}
\hspace*{-10mm}
\epsfxsize=140mm
\epsffile{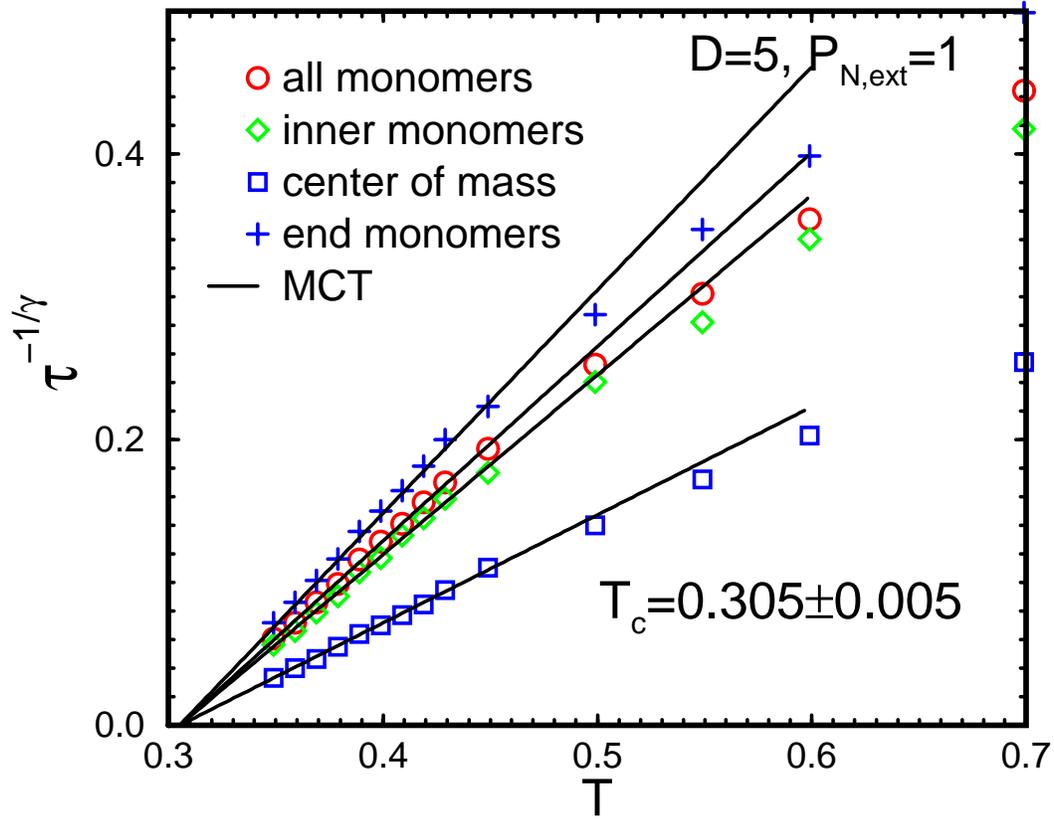}
\caption[]{Plot of $\tau^{-1/\gamma}$ versus $T$. 
The relaxation time $\tau$ was determined by 
Eq.~(\ref{eq:tau_gi::def}) using the mean-square displacements of
inner, end and all monomers and of the chain's center of mass.
The mode-coupling exponent $\gamma$ was first determined from fits to 
Eq.~(\ref{eq:MCT::power::law::for::tau})
where all fit parameters were first treated as independent.
As all values obtained for $\gamma$
agreed within the error bars, $\gamma \myeq 2.5\pm 0.2$, we 
repeated the fits with $\gamma \myeq 2.5$.
The MCT fits to the data are represented by the straight lines.
From the intersection of these lines with the $T$-axis the critical 
temperature is  determined $\Tc(D\myeq 5) \myeq 0.305 \pm 0.005$.
}
\label{fig:tau_g0_g4_eq0.667.D5_vs_T}
\end{figure}

Table ~\ref{tab1} contains results for $\Tc(D)$ and 
$\gamma(D)$ obtained in this way.
Figure~\ref{fig:tau_g0_g4_eq0.667.D5_vs_T} shows a representative
example for this analysis. It depicts $\tau^{-1/\gamma}$
versus $T$ for a film of thickness $D\myeq 5$
(note that $D$ is the wall-to-wall distance, the thickness of the region 
with non-vanishing monomer density is approximately $D\! - \!2$, see 
Fig.~\ref{fig:density_profiles_p1_D20_T_sym}).
The intersection of the straight lines (MCT-fit results) with the 
$T$-axis yields the critical temperature at this film thickness: 
$\Tc(D \myeq 5) \myeq 0.305 \pm 0.005$.
Note that, despite the highly non-linear relationship between
the MSD's used to define the  various $\tau$'s, all 
fits yield the same $\Tc$.

\begin{figure}
\hspace*{-23mm}
\epsfxsize=140mm 
\hspace*{20mm}
\epsffile{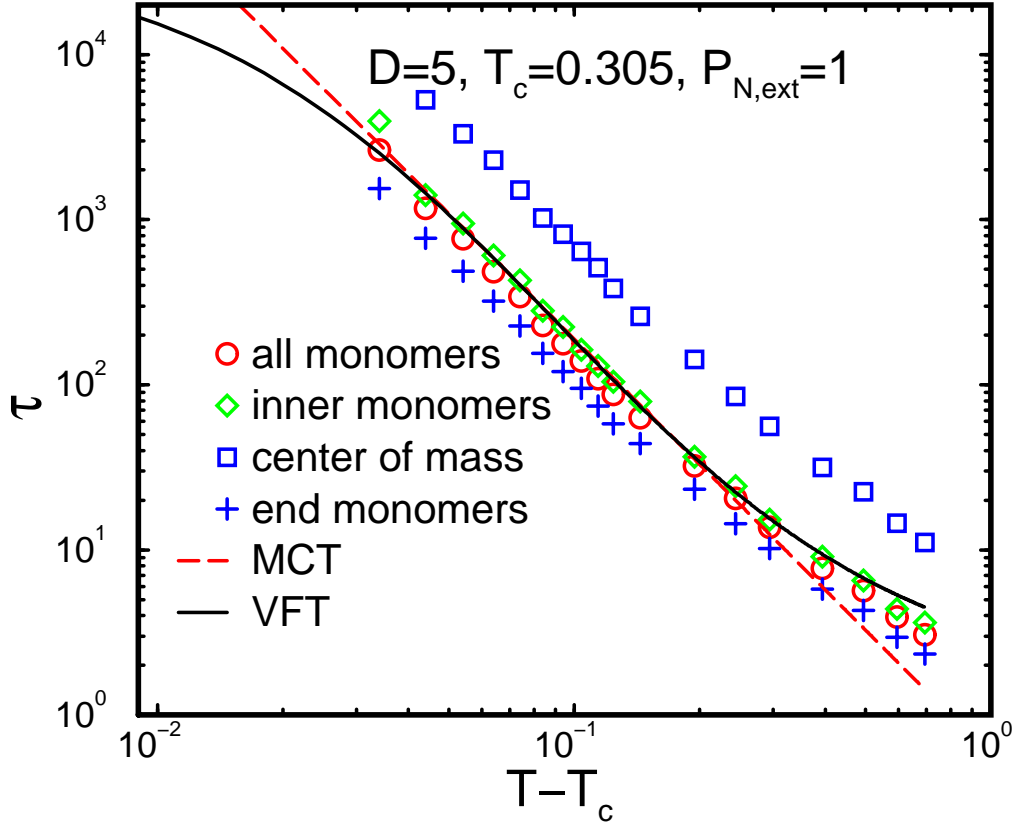}
\caption[]{Relaxation time $\tau(g_i\myeq 1)$ for an extremely 
thin film of three monomer layers only (wall-to-wall distance $D\myeq 5$).  
Different mean-square displacements are used for the analysis:
$g_0$ = MSD of all monomers, 
$g_1$ = MSD of the innermost monomer,
$g_3$ = MSD of the chain's center of mass and
$g_4$ = MSD of the end monomers.
The long-dashed line indicates the fit using 
Eq.~(\ref{eq:MCT::power::law::for::tau})
 motivated by the ideal mode-coupling theory.
However, the solid line which corresponds to a fit using the VFT-law
describes  equally well the data [see Eq.~(\ref{eq:VFT::law::for::tau})].
Both fits shown here were done for $g_1(t)$.
}
\label{fig:tau_g1_eq0.667.D5.compare.VFT_w_MCT}
\end{figure}

\begin{figure}
\hspace*{-23mm}
\epsfxsize=140mm 
\hspace*{20mm}
\epsffile{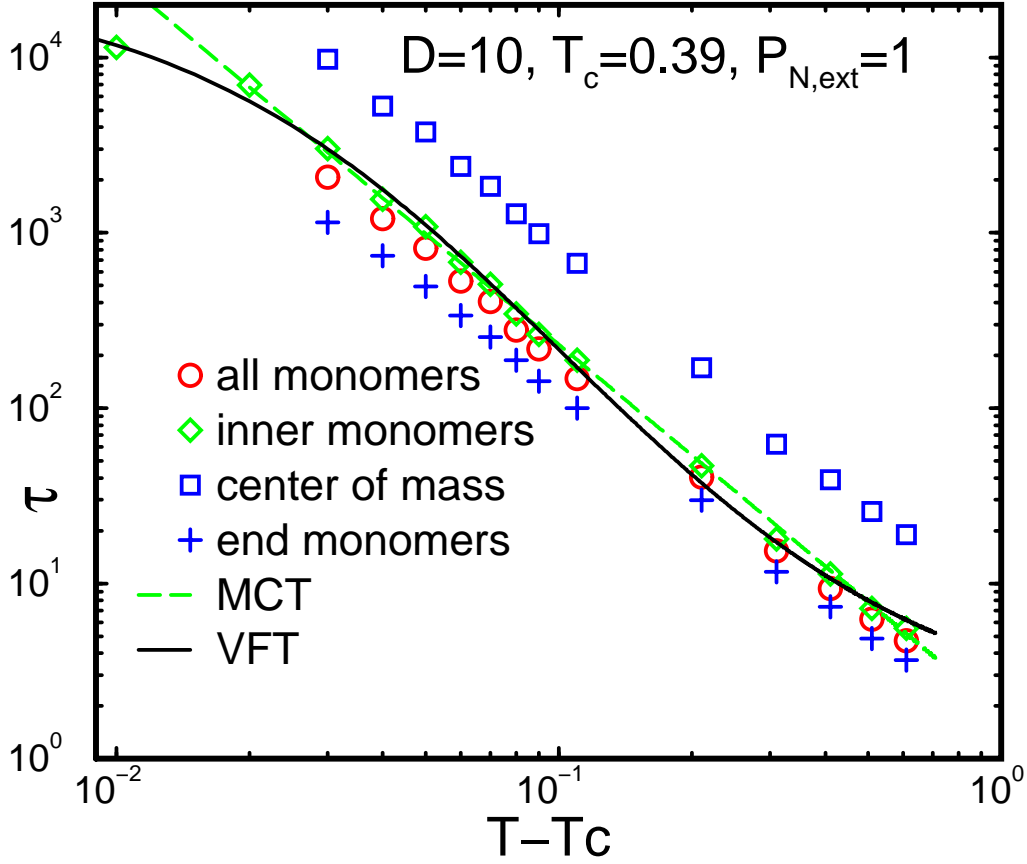}
\caption[]{Relaxation time $\tau(g_i\myeq 1)$ for a film
of thickness $D\myeq 10$.
Different mean-square displacements are used for the analysis:
$g_0$ = MSD of all monomers, 
$g_1$ = MSD of the innermost monomer,
$g_3$ = MSD of the chain's center of mass and
$g_4$ = MSD of the end monomers.
The long-dashed line indicates the fit using 
Eq.~(\ref{eq:MCT::power::law::for::tau})
 motivated by the ideal mode-coupling theory.
The solid line corresponds to a fit using the VFT-law 
[see Eq.~(\ref{eq:VFT::law::for::tau})].
Both fits shown here were done for $g_1(t)$.
}
\label{fig:tau_g1_eq0.667.D10.compare.VFT_w_MCT}
\end{figure}

To test this analysis the resulting critical temperature can be 
used to linearize reduced plots of the relaxation times 
versus $T\!-\!\Tc$ on a  log-log scale. 
Figures~\ref{fig:tau_g1_eq0.667.D5.compare.VFT_w_MCT} 
and~\ref{fig:tau_g1_eq0.667.D10.compare.VFT_w_MCT}
show that
the power law~(\ref{eq:MCT::power::law::for::tau}) motivated 
by the ideal MCT is a good approximation of the data at temperatures 
close (but not too close) to $\Tc$ (see, for example,
Fig.~\ref{fig:tau::g1::versus::T-Tc}
and also~\cite{Bennemann-Baschnagel-Paul::EPJB10,%
Aichele-Baschnagel::EurPhysJE::I,Bennemann-Baschnagel-Paul-Binder}
for comparable bulk data).

\begin{figure}
\epsfxsize=140mm
\hspace*{0mm}
\epsffile{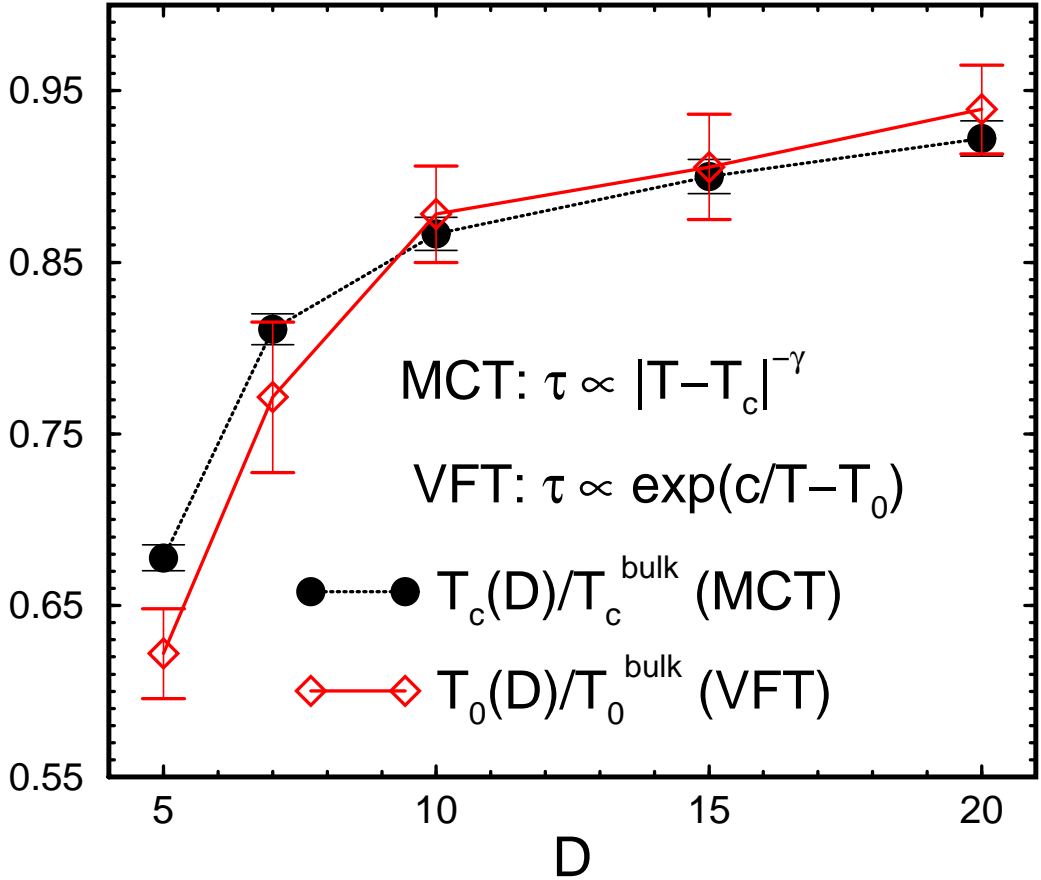}
%\epsfxsize=110mm
%\epsfysize=110mm
%\hspace*{15mm}\epsffile{TgKeddieJones.ps}
\caption[]{Ratios $T_\mr{c}(D)/T_\mr{c}^\mr{bulk}$
and $T_\mr{0}(D)/T_\mr{0}^\mr{bulk}$ versus film thickness $D$.
The critical temperatures, $T_\mr{c}(D)$, of the films 
were obtained from fits to Eq.~(\ref{eq:MCT::power::law::for::tau}). Similarly, the  
Vogel-Fulcher-Tammann temperatures, $\Tnull(D)$, and $T^{\rm bulk}_0$ are results of
fits to  Eq.~(\ref{eq:VFT::law::for::tau}). The error bars of $T_0(D)$ are
larger than those of $\Tc(D)$, since $T_0(D)\ll \Tc(D)$ so that the 
difference between the lowest simulated temperature and the 
extrapolated reslut is much larger for the VFT-temperature
than for the critical temperature.
Therefore, the results for $\Tc(D)$ are more reliable.
}
\label{fig:Tc_of_D}
\end{figure}

\begin{figure}
\epsfxsize=140mm
\hspace*{-7mm}
\epsffile{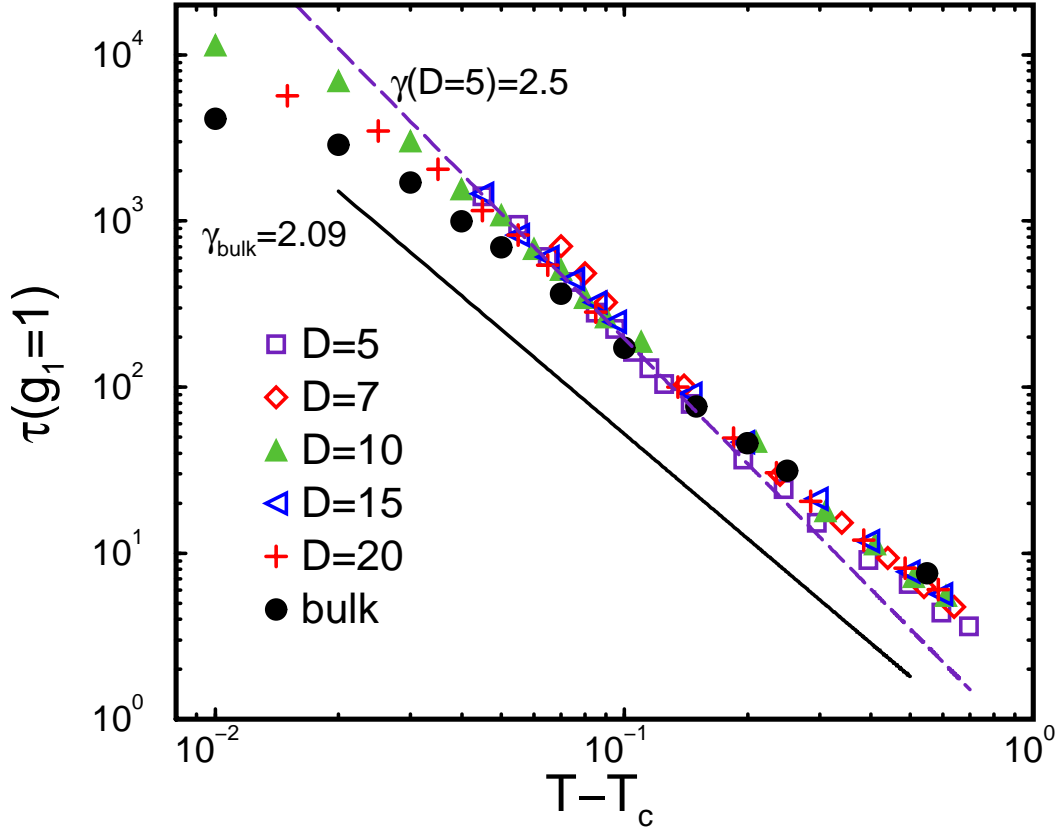}
\caption[]{Relaxation  time $\tau$ determined from $g_1(t)$ by requiring 
$g_1(\tau)\myeq 1$(= monomer diameter). The solid line indicates the power 
law $\tau \sim (T-T_\mr{c}^\mr{bulk})^{-\gamma_\mr{bulk}}$ suggested by MCT for the 
bulk at temperatures above $\Tcbulk$
($\Tcbulk \myeq 0.45$ and  $\gamma_\mr{bulk}
\myeq 2.09$~\cite{Bennemann-Baschnagel-Paul::EPJB10}).
The long-dashed line represents a fit to $D\myeq 5$ data
with an exponent of $\gamma(D\myeq 5)\myeq 2.5$.
Both in the film and in the bulk, for temperatures very 
close to $\Tc$, the relaxation times increase more slowly than
predicted by the ideal MCT.
}
\label{fig:tau::g1::versus::T-Tc}
\end{figure}

However, as indicated by the solid line in the both figures,
$\tau(T)$ can also be described by a Vogel-Fulcher-Tammann (VFT)-fit 
in the studied temperature range, i.e. by
\begin{equation}
\tau(T) \propto  \exp \Big( \myfrac{c(D)}{T-\Tnull(D)} \Big) \; ,
\label{eq:VFT::law::for::tau}
\end{equation}
where $c$ is a  constant which can depend on film thickness.
The possibility of describing the same data both by a power law (MCT)
and by a VFT-fit has also been observed for bulk properties of the 
present model (see Fig. 10 in~\cite{Bennemann-Paul-Binder-Duenweg::PRE57}).
We therefore use the VFT-formula as an independent approach and determine 
the VFT-temperature, $\Tnull$, for various film thicknesses. 
Table~\ref{tab1} contains the results for $T_0(D)$ thus obtained.
A plot of $\Tc(D)/\Tcbulk$ and 
$\Tnull(D)/\Tnullbulk$ is shown in
Fig.~\ref{fig:Tc_of_D}. Both the mode-coupling critical temperature
and the VFT-temperature exhibit similar $D$-dependences, thus 
suggesting that the presence of the smooth walls results
in a reduction of the glass transition temperature, $\Tg$.
As already mentioned in section~\ref{section::introduction}, a 
reduction of the glass transition temperatures has also 
been reported both from experiments on supported polymer 
films~\cite{Keddie::Jones::Cory::EuroPhysLett28} and on
free standing polymer
films~\cite{Forrest::2000,Dalnoki-VeressMurray,Forrest::PRL77::page2002}.
and also from MD simulations of a model of square-well spherical interaction 
sites interconnected by fully flexible strings
in the case of a weak attraction between the substrate and the 
chains~\cite{Torres-Nealey-dePablo::PRL85}.

\begin{figure}
\epsfxsize=140mm
\hspace*{0mm}
\epsffile{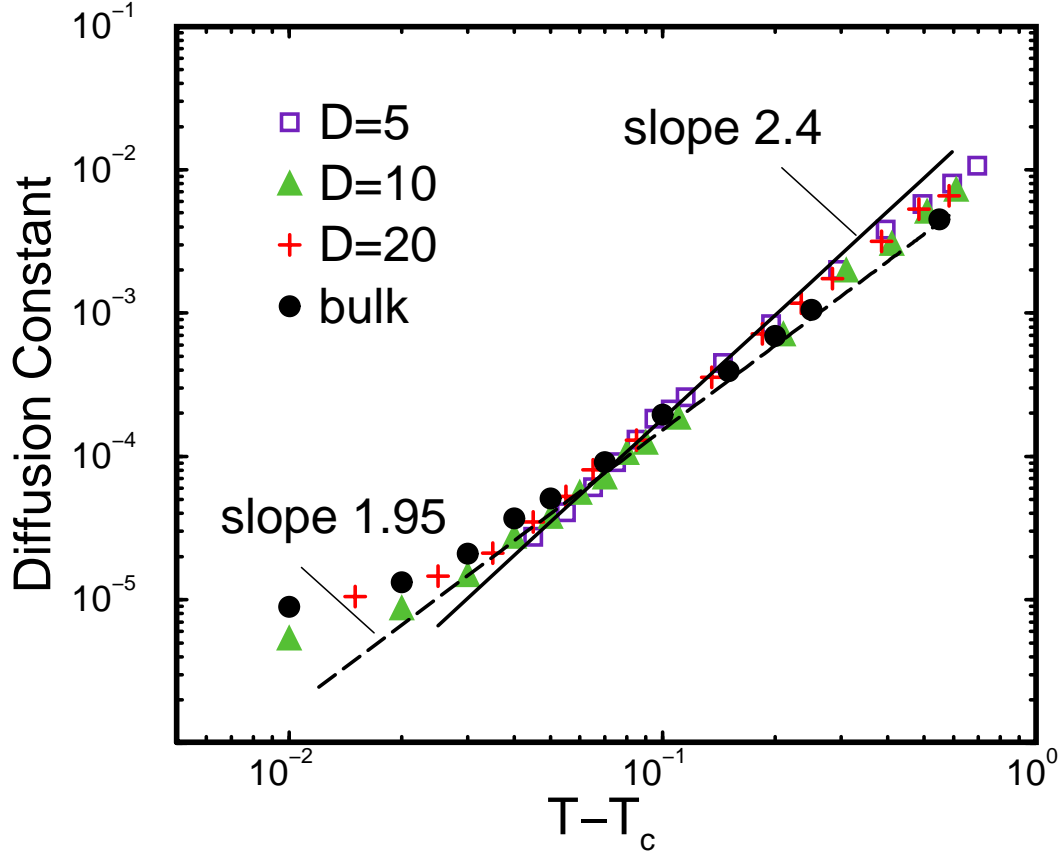}
\caption[]{Diffusion constant of a chain versus $T-\Tc$ for films of 
thickness $D\myeq5,\; 10,\; 20$
and for the bulk. The critical temperatures, $T_\mr{c}(D)$, of the films 
were obtained from fits to 
Eq.~(\ref{eq:MCT::power::law::for::tau}) [see table~\ref{tab1}].
$T^{\rm bulk}_{\rm c}$
was known from the previous bulk 
analysis~\cite{Bennemann-Baschnagel-Paul::EPJB10}.
The straight lines indicate fits using 
Eq.~(\ref{eq:MCT::power::law::for::diffconst}).
The mode-coupling exponent $\gamma$ is larger for stronger confinement 
(smaller film thickness).
Note that, the lowest simulated temperature is $T\myeq 0.35$ for $D\myeq 5$ 
and  $T\myeq 0.4$ for $D\myeq 10$. However, as the critical temperature 
of the thinner film
is much lower ($\Tc(D\myeq 5)\myeq 0.305$ compared to $\Tc(D\myeq 10)\myeq 0.39$)
the $D\myeq 10$ data are {\em closer} to the corresponding $\Tc$. This 
explains why practically no deviation
from a power low is observed for $D \myeq 5$.
}
\label{fig:diffconst::vs::T-Tc}
\end{figure}

\begin{figure}
\epsfxsize=140mm
\hspace*{0mm}
\epsffile{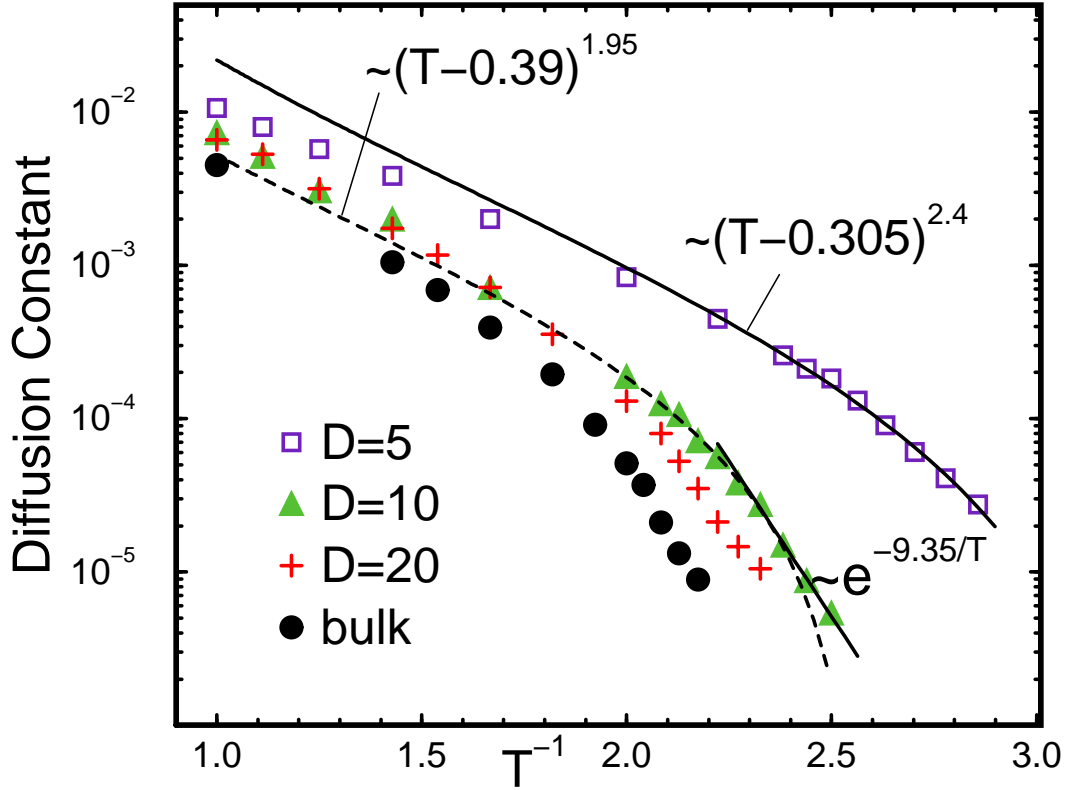}
\caption[]{
Diffusion constant of a chain versus inverse temperature for films of 
thickness $D\myeq5,\; 10,\; 20$
and for the bulk. The critical temperatures, $T_\mr{c}(D)$, of 
the films were obtained from fits to
Eq.~(\ref{eq:MCT::power::law::for::tau}) [see table~\ref{tab1}].
$T^{\rm bulk}_{\rm c}$ was known from the previous bulk 
analysis~\cite{Bennemann-Baschnagel-Paul::EPJB10}.
The solid line indicates the fit results using 
Eq.~(\ref{eq:MCT::power::law::for::diffconst}) 
for $D \myeq 5$. The dashed line indicates a similar fit 
result for $D\myeq 10$.
A short solid line indicates the result of an Arrhenius 
fit [see Eq.~(\ref{eq:diffusion::Arrhenius})] to 
the $D\myeq 10$ data in a low temperature range, 
where the ideal MCT no longer holds [see also the text].
Note that, the lowest simulated temperature is $T\myeq 0.35$ for $D\myeq 5$ 
and  $T\myeq 0.4$ for $D\myeq 10$. However, as the critical 
temperature of the thinner film is much lower 
($\Tc(D\myeq 5)\myeq 0.305$ compared to $\Tc(D\myeq 10)\myeq 0.39$), 
the $D\myeq 5$ data are {\em farther} from the corresponding $\Tc$.
This explains why a power low fits well to $D\myeq 5$ 
data at all temperatures.
}
\label{fig:diffconst::vs::Tinv}
\end{figure}

In Fig.~\ref{fig:tau::g1::versus::T-Tc}, $\tau(g_1\myeq 1)$
is depicted  versus $T \! - \! \Tc$ for the homogeneous 
melt (bulk) and films of various thicknesses.
The solid line indicates the power
law $\tau \sim (T - \Tcbulk )^{-\gamma_{\rm bulk}}$ with $\Tcbulk \myeq 0.45$
and $\gamma_{\rm bulk} \myeq 2.09$~\cite{Bennemann-Baschnagel-Paul::EPJB10,%
Aichele-Baschnagel::EurPhysJE::I,%
Aichele-Baschnagel::EurPhysJE::II}. To show the (slight) $D$-dependence
of the exponent $\gamma$, a fit to $D\myeq 5$ data is also 
depicted (see the long-dashed line with an slope of 2.5).
It is seen from this figure that both in the film and in the bulk, 
for temperatures very close to $\Tc$, the relaxation times increase 
more slowly than predicted by the ideal MCT.  This discrepancy is an indication 
of slow relaxation processes which are not taken into account within the
ideal MCT. As temperature decreases, the contribution of these processes
becomes important and the ideal MCT no longer holds. A common picture
in describing the slow dynamics of the system at these temperatures
is that of an energy landscape with {\em finite} potential barriers.
Relaxation processes are then described as jumps (hoppings) between 
neighboring energy minima. If one assumes a sharp distribution of
the potential barriers around a given value, $E_0$, i.e. if all
such barriers are assumed to have more or less  the same height,
the probability of a jump over a barrier is then equal to 
the probability of having an energy $E \ge E_0$. 
Assuming that the diffusion constant 
is proportional to this probability, one obtains,
\begin{eqnarray}
{\rm Diffusion~Constant} & \propto &
\int^{\infty}_{E_0}
\exp(-\myfrac{E}{\kB T})\, d E = \exp(-\myfrac{E_0}{\kB T})\; .
\label{eq:diffusion::Arrhenius}
\end{eqnarray}
Thus, due to the above picture, at low enough temperatures
the diffusion constant obeys an Arrhenius law. Here, ``low enough'' 
means temperatures very close to or below $\Tc$. However, for higher 
(but not too high) temperatures one again expects the validity of 
a power law also for the diffusion constant,
\begin{equation}
\mbox{Diffusion~Constant} \propto |T-\Tc(D)|^{\gamma(D)}\; .
\label{eq:MCT::power::law::for::diffconst}
\end{equation}
Using the mean square displacements, we compute the diffusion 
constant from
\begin{equation}
{\rm Diffusion~Constant}  =  \lim_{t \to \infty}  \myfrac{ g_{i} (t) }{6 t} \; .
\label{eq:diffusion::constant::film}
\end{equation}
Note that, in the diffusive limit ($t \to \infty$), there is no 
difference between $g_{0}$, $g_{1}$ , $g_{3}$ or $g_{4}$ 
(see for example the upper panel in Fig.~\ref{fig:g1g1.for_diff_D_and_T}, 
where $g_{0}(t)$ and $g_{3}(t)$ coinside for large $t$).
In the praxis, however, one has to evaluate Eq.~(\ref{eq:diffusion::constant::film})
at large but finite $t$. As the mean-square displacements of chain's center of mass
reaches the diffusive limit faster than other MSD's
(see Fig.~\ref{fig:g1.T0.48.bulk}), one should use $g_3(t)$ 
in an evaluation of Eq.~(\ref{eq:diffusion::constant::film}).

best results on the diffusion constant
one obtains it is   numerically 
A log-log plot of the diffusion constant versus $T\! - \!\Tc$
is depicted in Fig.~\ref{fig:diffconst::vs::T-Tc}
for various film thicknesses and for the bulk.
Similar to the behavior of the  relaxation time discussed above,
at temperatures very close to $\Tc$,  results for the diffusion 
constant deviate from the power law given by
Eq.~(\ref{eq:MCT::power::law::for::diffconst}). Motivated by the
discussion which led to Eq.~(\ref{eq:diffusion::Arrhenius}), we try to fit
the diffusion constant at law temperatures by an Arrhenius law.
This is demonstrated in Fig.~\ref{fig:diffconst::vs::Tinv} for the case 
of a film of thickness $D \myeq 10$, where Eq.~(\ref{eq:diffusion::Arrhenius})
is applied to the last few data points in low temperature regime.
From this fit, we obtain $E_0 \myapprox 9.35$.
However, one should be carefull with an interpretation of $E_0$.
Recall that Eq.~(\ref{eq:diffusion::Arrhenius}) was derived assuming 
that all potential barriers have the same height of $E_0$.
Therefore, an interpretation of $E_0$ as the typical height of 
the potential barriers in an energy landscape requires a study of
the inherent structure of the system.

It is also shown in Fig.~\ref{fig:diffconst::vs::Tinv} 
that the data at higher but not too high temperatures are
well described by a power law, as expected from 
the MCT analysis of the relaxation times 
presented in this section.

Next we look at the temperature dependence of a crossover 
time, $\tau_{\rm co}$, given by
\begin{eqnarray}
g_{1} (t \myeq \tau_{\rm co}) &:=& 6r^2_{\rm sc}=0.054
\mbox{~(definition of $\tau_{\rm co}$)}\; .
\label{eq:tau_co::def::bulk}
\end{eqnarray}
This definition is motivated by an analysis of the bulk, where 
$r_{\rm sc}$ ($\myapprox 0.1$ monomer diameter) appears as a 
relevant length scale for the dynamics of a monomer in its local
environment (``cage'')~\cite{Bennemann-Baschnagel-Paul::EPJB10,%
Aichele-Baschnagel::EurPhysJE::I}. 
The value $6r^2_\mr{sc}$ (roughly) coincides with the inflection point
of $g_1(t)$ and thus marks the crossover from the early
time regime, where $g_1(t)$ is almost flat, to the late time regime, 
where $g_1(t)$  gradually increases with time and finally becomes  diffusive 
(see Fig.~\ref{fig:g1g1.for_diff_D_and_T}).  Motivated by ideal MCT, 
one expects that $\tau_{\rm co}$ diverges at $\Tc$ with a power law,
\begin{equation}
\tau_{\rm{co}}(T) \propto |T-\Tc|^{\myfrac{-1}{2a}}\; ,
\label{eq:MCT::power::law::for::tau::crossover}
\end{equation}
with $0\!<\!a\!<\!0.5$. The exponent 
$a$ describes the initial decay of the correlation 
functions from the ballistic regime to the plateau~\cite{Goetze:Liquids::1991,%
Goetze-Sjoegren::TransportTheoryStatPhys,%
Goetze-Sjoegren::RepProgPhys55}.
The analysis of the bulk data showd that 
$a\myeq 0.352$~\cite{Bennemann-Baschnagel-Paul::EPJB10}.

\begin{figure}
\epsfxsize=130mm
\hspace*{-5mm}
\epsffile{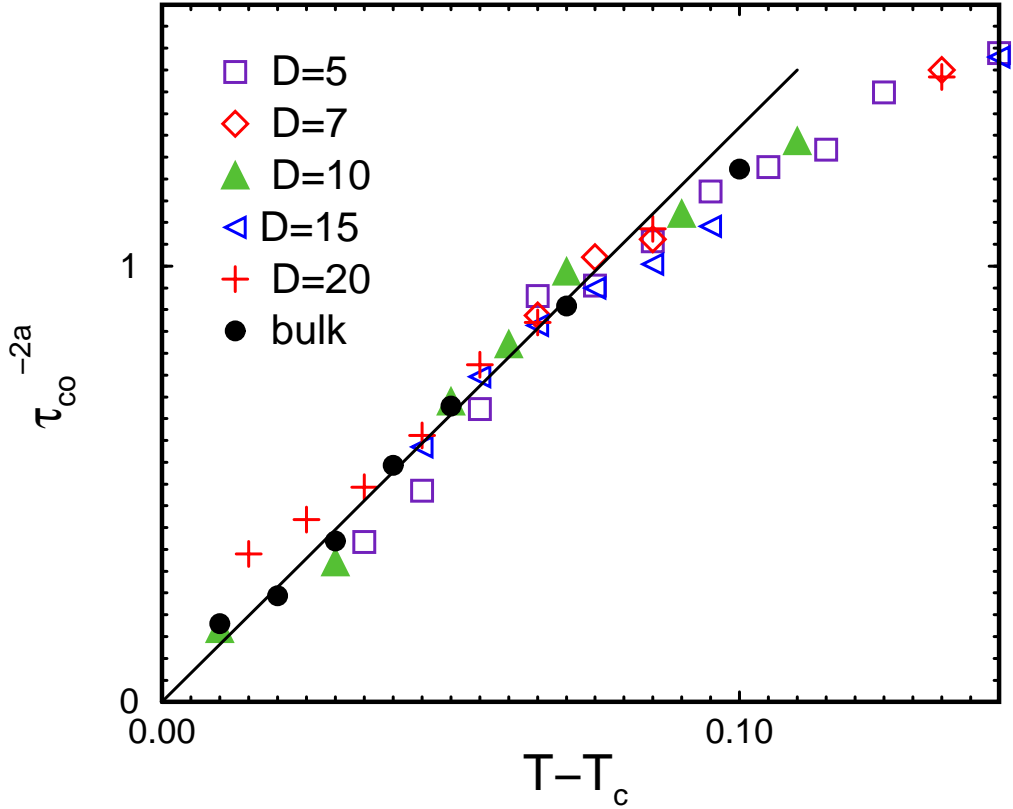}
\caption[]{Temperature dependence of the crossover time $\tau_\mr{co}$ and the 
relaxation time $\tau(g_1\myeq 1)$ (lower panel) for the bulk and different 
film thicknesses: $D\myeq 5$ ($\approx \! 3.5 R_\mr{g}$; $R_\mr{g}\myeq $ 
bulk radius of gyration), $D\myeq 7$ ($\approx \! 5 R_\mr{g}$), 
$D\myeq 10$ ($\approx \! 7 R_\mr{g}$) and $D\myeq 20$ ($\approx \! 
14 R_\mr{g}$). The critical temperatures
were obtained from fits to Eq.~(\ref{eq:MCT::power::law::for::tau}) 
[see Table~\ref{tab1}]. The crossover time was determined from the 
mean-square displacement of the innermost 
monomer, $g_1(t)$, by the condition 
$g_1(\tau_\mr{co}) \myeq  6r^2_\mr{sc} \myeq  0.054$. This choice as 
well as power laws of both panels are
motivated by a MCT-analysis of the bulk data
\cite{Bennemann-Baschnagel-Paul::EPJB10,%
Aichele-Baschnagel::EurPhysJE::I,%
Aichele-Baschnagel::EurPhysJE::II}. 
The value $6r^2_\mr{sc}$
(roughly) coincides with the inflection point of $g_1(t)$ and 
thus marks the crossover from the early
time regime, where $g_1(t)$ is almost flat, to the late time 
regime, where $g_1(t)$ increases faster
with $t$ and finally becomes diffusive [see Fig.~\ref{fig:g1g1.for_diff_D_and_T}]. 
The exponent $a$ 
used to linearize the data close to $T_\mr{c}$ is that of the bulk 
($a\myeq 0.352$~\cite{Bennemann-Baschnagel-Paul::EPJB10}).
}
\label{fig:tau::crossover}
\end{figure}

\begin{figure}
\epsfxsize=100mm
\hspace*{15mm}\epsffile{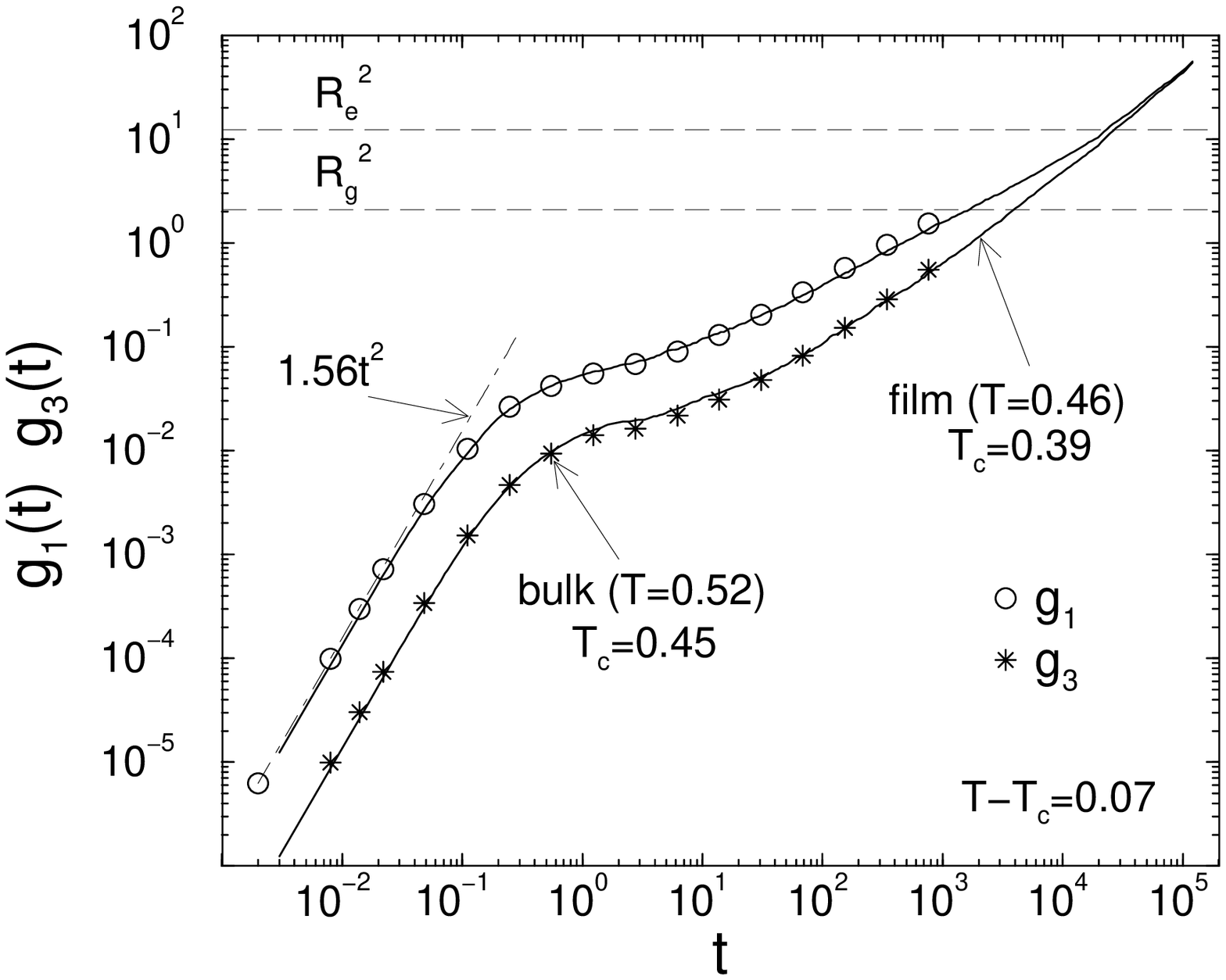}
\epsfxsize=100mm
\hspace*{21mm}\epsffile{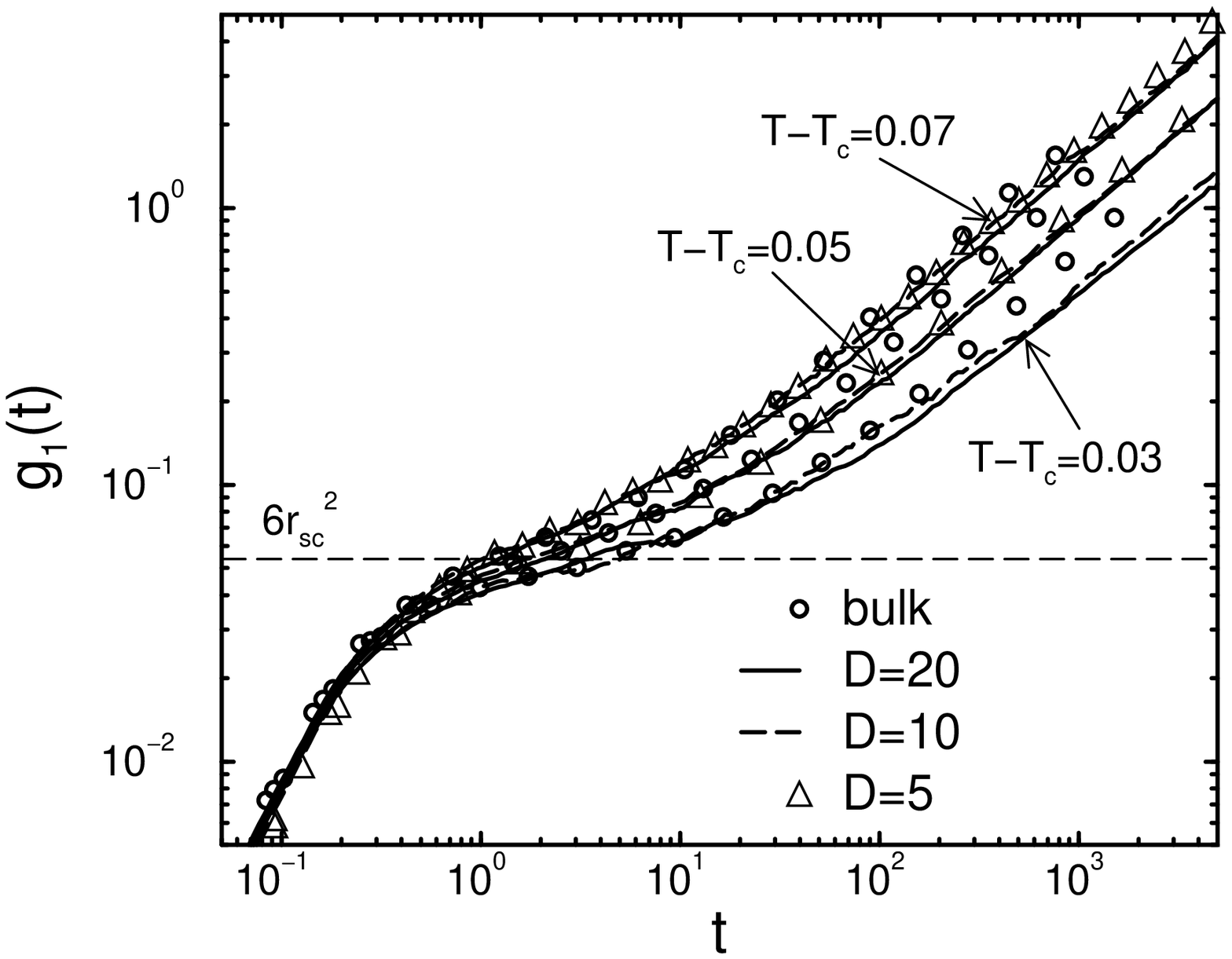}
\caption[]{The upper panel shows the mean-square displacements of the 
innermost monomer, $g_1(t)$, 
and of the chain's center of mass, $g_3(t)$, between the bulk (symbols 
$\circ$ and $\ast$) and a 
film (solid lines) of thickness $D\myeq 10$ ($\approx \! 7 R_\mr{g}$). 
The panel compares the results for 
the same distance to the critical temperature $T_\mr{c}$ (i.e., 
$T-T_\mr{c}\myeq 0.07$). The displacements 
of the film are calculated parallel to the wall. They were multiplied 
by $3/2$ to put them on the 
scale of the bulk data. The two horizontal dashed lines show the 
bulk end-to-end distance 
$R^2_\mr{e}$ and the radius of gyration $R_\mr{g}^2$, respectively. 
The ballistic short-time
behavior $g_1(t)\myeq 3Tt^2$ is indicated for the bulk at $T\myeq 0.52$. 
The lower panel shows
behavior of $g_1(t)$ for $D\myeq 5,\; 10,\; 20$ and for the bulk 
when approaching $T_\mr{c}$ ($T_\mr{c}(D\myeq 5)\myeq 0.305$,
$T_\mr{c}(D\myeq 10)\myeq 0.39$, 
$T_\mr{c}(D\myeq 20)\myeq 0.415$, 
$T_\mr{c}^\mr{bulk}\myeq 0.45$).
The dashed horizontal line shows the plateau value $6r_{\rm sc}^2$ 
($\simeq 0.054$)
of a MCT-analysis performed on the bulk data 
\cite{Bennemann-Baschnagel-Paul::EPJB10}.
For intermediate times, the film and bulk data 
corresponding to a given $T-\Tc$ follow the same master curve.
}
\label{fig:g1g1.for_diff_D_and_T}
\end{figure}

Figure~\ref{fig:tau::crossover} displays 
$\tau^{-2a}_\mr{co}$ versus $T\! - \! \Tc$
for the bulk and for various film thicknesses: $D\myeq 5$ 
($\approx \! 3.5 R_\mr{g}$; $R_\mr{g}\myeq$
bulk radius of gyration), $D\myeq 7$ ($\approx\! 5 R_\mr{g}$), 
$D\myeq 10$ ($\approx\! 7 R_\mr{g}$) 
and $D\myeq 20$ ($\approx\! 14 R_\mr{g}$).
The critical temperatures are taken from 
Table~\ref{tab1} and for the exponent $a$ we used the bulk value
$a\myeq 0.352$~\cite{Bennemann-Baschnagel-Paul::EPJB10}. 
It is seen from Fig.~\ref{fig:tau::crossover} 
that this value yields a good linearization of the bulk data.
For $\tau_\mr{co}$ obtained from the film data, however, 
deviations occur indicating a (rather slight) $D$-dependence of
the exponent $a$.

The previous analysis of the $T$-dependence of the relaxation time
$\tau$ and of the crossover time $\tau_{\rm co}$ suggests that
the critical temperature depends strongly on film thickness,
but the variation of the exponents $a$ or $\gamma$ is rather weak.
Within the framework of MCT the exponents $a$ and $\gamma$ are
related to one another and determined by the so-called
exponent parameter $\lambda$, which controls the time 
evolution of the monomer mean-square displacement in 
the plateau regime (see Fig.~\ref{fig:g1.T0.48.bulk}).
Thus, the weak variation of the exponents $a$ and $\gamma$
with $D$ should imply that the time dependence of the
mean-square displacements in the film closely follows
that of the bulk, in the plateau regime. It is therefore 
interesting to test whether the mean-square displacements of
the film and of the bulk obey the same master curve when
compared for the same reduced temperature, $T\!-\!\Tc$.
This idea is first examined in the upper panel of
Fig.~\ref{fig:g1g1.for_diff_D_and_T} where the MSD's of the 
innermost monomers, $g_1(t)$,
and of the chain's center of mass, $g_3(t)$,  
are compared at $T-\Tc \myeq 0.07$
for the bulk and for a film of thickness $D\myeq 10$
(as $T^{\rm bulk}_{\rm c}\myeq 0.45$ and $\Tc(D\myeq 10)\myeq 0.39$,
the bulk data at $T\myeq 0.52$ are compared to the corresponding film
data at $T\myeq 0.46$).
There is a good agreement between film and bulk data over a 
long period of time, in particular for intermediate times.
The lower panel of the figure extends this comparison to 
lower temperatures. Whereas the film data still 
closely agree with each other for all times simulated, this is 
no longer the case for the bulk data. By simply shifting the
$T$-axis it is only possible to rescale bulk and film data
onto a common master curve in the intermediate time window of 
the plateau. At later times, however, $g_1(t)$ of the bulk
increases faster with $t$ than $g_1(t)$ of the films. 
This difference is more  pronounced the 
closer $T$ to $\Tc(D)$.

A preliminary analysis of the incoherent scattering function 
also reveals the property that for intermediate times, 
the data corresponding to different film thicknesses follow 
the same master curve when compared for the same reduced 
temperature $T-\Tc(D)$. A more detailed investigation and comparison 
with the behavior of the mean-square displacements is under way.

%%%%%%%%%%%%%%%%%%%%%%%%%%%%%%%%%%%%%%%%%%%%%%%%%%%%%%%%%%%%%%%%%%%%
%%%%%%%%%%%
%%%%%%%%%%%
\section{Local Dynamics}
\label{section:Local_Dynamics}
\begin{figure}
\epsfxsize=130mm
\hspace*{0mm}
\hspace*{5mm}
\epsffile{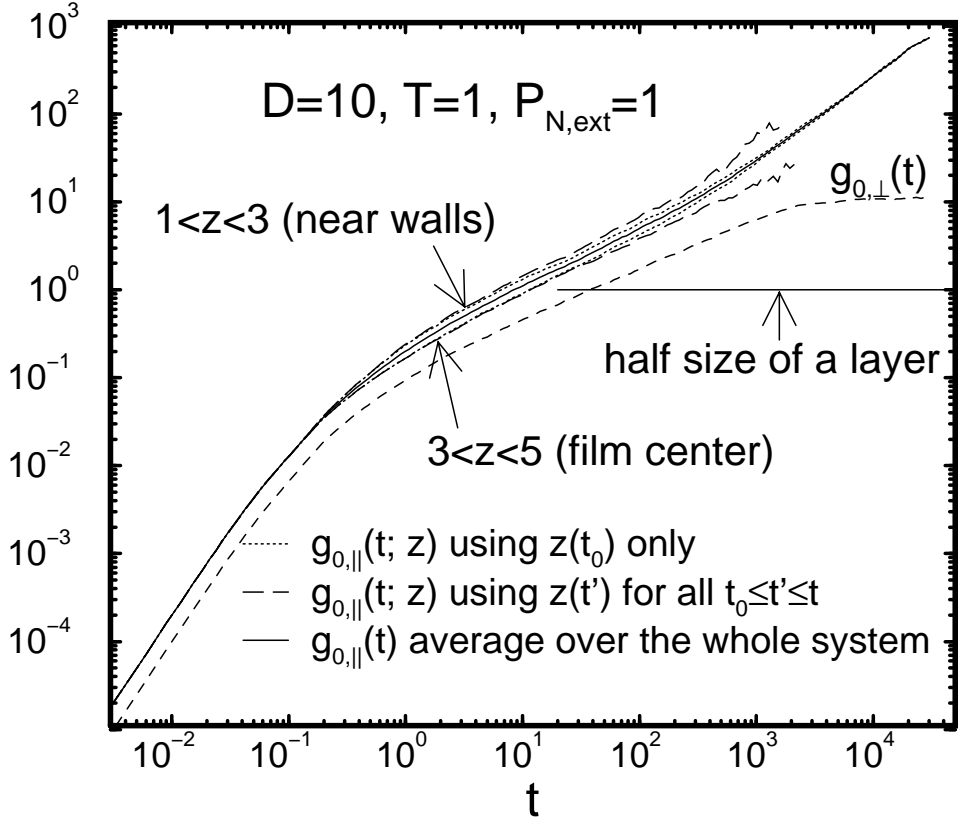}
\caption[]{Illustration of the dependence of the local MSD upon the way the 
particle displacements are attributed to layers (``localized'').
For this purpose, two regions of different mobility are considered:
A region around the film center, $3 \le z \le 5$, and the first layer 
close to the walls, $1\le z\le 3$ ($z\myeq |z_{\rm particle}-z_{\rm wall}|$).
Note that there is practically no particle  in the very proximity of the
walls, i.e. in the region with $0\le z\le 1$.
If the displacement of a particle is allways attributed to 
the layer, where the particle was observed at $t_0$,
the obtained results for the local MSD's are biased in 
the following sense: If a tagged particle leaves its initial layer 
at a later time, its motion will partly represent 
the properties of the layers it has ``visited'' so far.
Attributing the mobility of such a  particle to the initial 
layer thus represents an averaging over many layers. The error 
introduced in this way is negligible for short times. 
However, as the MSD in transversel 
direction, $g_{0,\bot}(t)$, becomes comparable to the half of the layer 
thickness [see the intersection of the horizontal line with $g_{0,\bot}(t)$]
the error dominates and the local character of the data is 
lost to a large extent. For very long times, no local 
information ``persists'' this averaging and curves belonging to 
different regions converge towards the system average [see long time 
behavior of the dotted lines].
The correct definition requires that only those particle are allowed to 
contribute to the mobility of a given region, which stay
in the same layer for all times $t_0 \le t' \le t$ [see long-dashed lines].
However, there is also a drawback of this correct definition:
As any particle which leaves its initial layer must be excluded by the
computation of the MSD's, the statistical uncertanity increases with the
number of such particles and thus with time 
[see long-dashed lines for large $t$]. Statistical accuracy  at 
large times thus requires more independent samples of the same system.

The difference in the magnitude of $g_{0,\bot}(t)$ and 
$g_{0,\parallel}(t)$ arises from the fact that 
there is only one independent direction perpendicular 
to the walls compared to two independent parallel directions.}
\label{fig:displacements_xy_T1_p1_D10_compare_localization_definitions}
\end{figure}

%%%%%%%%%%%
%%%%%%%%%%%
%%%%%%%%%%%%%%%%%%%%%%%%%%%%%%%%%%%%%%%%%%%%%%%%%%%%%%%%%%%%%%%%%%%%
The discussion of the previous section illustrated that the
dynamics of the film is accelerated with respect to the bulk.
Obviously, this is a consequence of the walls. A natural 
question is therefore how the motion of the monomers 
depends on their distance from the wall. To obtain a better 
insight into this dependence we study
the local displacements, i.e.\ the mean-square displacements 
measured within layers of a small thickness (``bin'') at various distances 
from the wall. However, when trying to analyse local displacements the 
problem of how to assign particles to layers has to be addressed.
A possible definition is to associate a monomer labeled $i$ 
to that bin to which its initial $z$-position, $z_i(t_0)$, 
corresponds,
\begin{equation}
g_{0}(t;z) = \Big< \sum_{i=1}^{N} \delta(z-z_i(t_0))\, 
(\mbf{r}_i(t)-\mbf{r}_i(t_0))^2 /\sum_{1}^{N}\delta(z-z_i(t_0)) \Big>\; .
\label{eq:local::MSD::wrong::def}
\end{equation}
The drawback of this definition is, however, that some
particles may leave this layer  at a later time. If one 
adds their contribution to the MSD to that of those particles 
which allways remain in the initial region,
a sort of averaging over other regions is introduced.
Obviously, the larger the time
difference $t-t_0$, the greater the probability
that some tagged particle has left its initial layer.
As a consequence, the local character of the obtained 
information becomes questionable for large times.
An estimate of the time beyond which
this averaging over neighboring regions becomes 
apreciable is the time at which the mean-square 
displacements in transversel direction become comparable 
to the half of the bin size.
For larger times local information 
is gradually lost and the MSD's of different 
regions converge towards the average MSD of the whole system.
This point is nicely demonstrated in
Fig.~\ref{fig:displacements_xy_T1_p1_D10_compare_localization_definitions}
where the MSD of all monomers in the direction parallel to the walls,
$g_{0,\parallel}(t)$, is shown for two regions of different mobility 
in a film of thickness $D\myeq 10$ at $T\myeq 1$ (high temperature liquid state).
As $g_{0,\parallel}\!>\!1$ (half of the bin size),
the MSD of both regions converge towards that of the 
whole system for large times.

To avoid this problem, a more stringent criterion must be
used to compute local  MSD's from particle displacements:
The contribution of a tagged particle to the MSD of a given layer
at a time $t$ should be taken into account if and only if the 
tagged particle has been in the same layer for 
{\em all times} $t_0 \le t' \le t$,
\begin{equation}
g_{0}(t;z) = \Big< \sum_{i=1}^{N} \prod_{t'=t_0}^{t} \delta(z-z_i(t'))\,
(\mbf{r}_i(t)-\mbf{r}_i(t_0))^2 /\sum_{1}^{N}\prod_{t'=t_0}^{t} 
\delta(z-z_i(t')) \Big> \; .
\label{eq:local::MSD::correct::def}
\end{equation}
This correct definition is, however, computationally mor demanding.
The reason lies in the fact that the contibution 
of particles which have left their initial layer at some later time, 
has to be ignored for all  subsequent times. As the number of such 
particles increases with time, the statistical accuracy decreases at 
late times. To improve it, one needs more independent samples of the same system.

\begin{figure}
\epsfxsize=140mm
\hspace*{0mm}
\epsffile{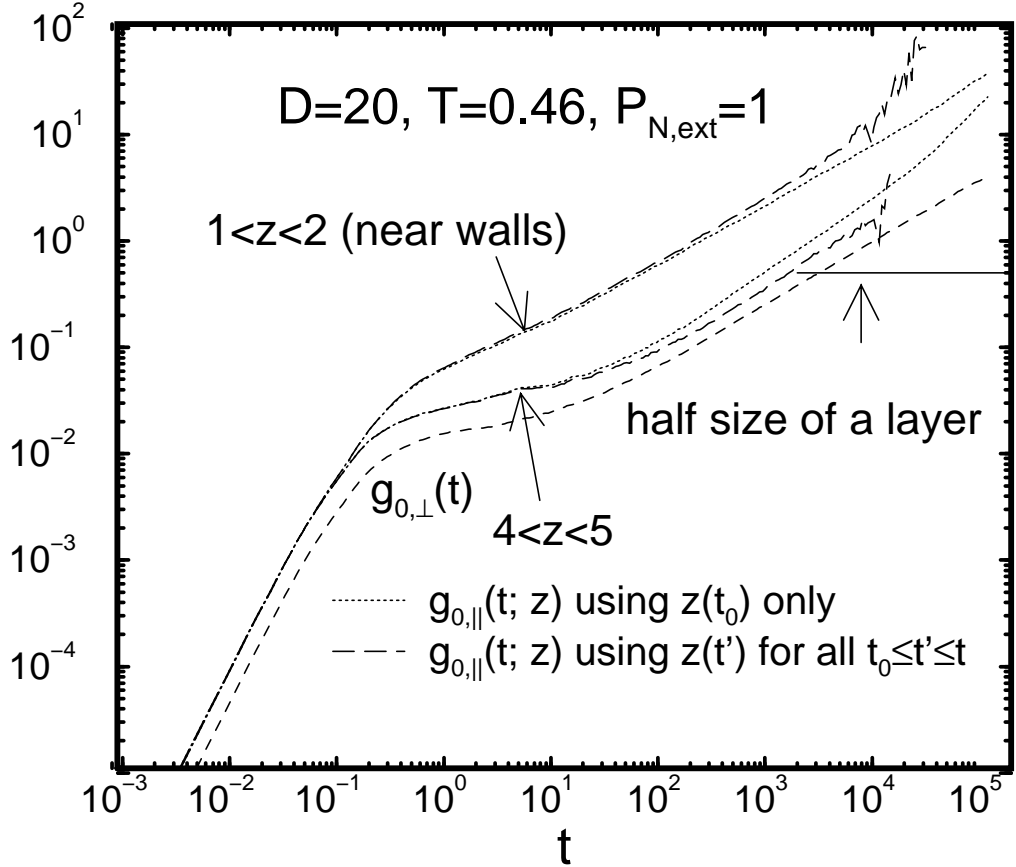}
\caption[]{Same as in 
Fig.~\ref{fig:displacements_xy_T1_p1_D10_compare_localization_definitions}
but for a film of thickness $D\myeq 20$ at a  low temperature of 
$T\myeq 0.46$ [$\Tc(D\myeq 20) \myeq 0.415$]. 
The figure illustrates that attributing the displacement of a 
particle to the layer, where it was at the initial time $t_0$,
can lead to errors which strongly depend on the layer
under consideration. Therefore, the results obtained using
this definition are not reliable even at times for which 
the MSD in transverse direction, 
$g_{0,\bot}$, is much smaller than the half of the layer 
thickness which, in turn, is the localization resolution.
The difference in the magnitude of $g_{0,\bot}(t)$ and 
$g_{0,\parallel}(t)$ arises from the fact that 
there is only one independent direction perpendicular 
to the walls compared to two independent parallel directions.}
\label{fig:displacements_xy_T0.46_p1_D20_compare_localization_definitions}
\end{figure}

In Fig.~\ref{fig:displacements_xy_T0.46_p1_D20_compare_localization_definitions}
we focus on another aspect of the same problem.
Figure~\ref{fig:displacements_xy_T1_p1_D10_compare_localization_definitions}
suggests that a good estimate of the time at which the difference between 
Eqs.~(\ref{eq:local::MSD::wrong::def}) and~(\ref{eq:local::MSD::correct::def}) 
is no longer negligible is $g_{1,\parallel}(t) \myeq \mbox{bin width}/2$.
Here, we stress that the inverse is not necessarily true,
i.e.\ the error is not necessarily negligible for shorter times. 
Already in 
Fig.~\ref{fig:displacements_xy_T1_p1_D10_compare_localization_definitions}
one can observe that, in the case of the layer close to the wall,
deviations from the correct curve occcur at shorter times than
the intersection point between $g_{0,\bot}(t)$ and the horizontal 
line indicating  half of the layer thickness.
This point is more clearly demonstrated in
Fig.~\ref{fig:displacements_xy_T0.46_p1_D20_compare_localization_definitions}
where the MSD's of a film of thickness $D\myeq 20$ at a low temperature
of $T\myeq 0.46$ are used [$\Tc(D\myeq 20)\myeq 0.415$].
Here, for the layer in the inner part of the system, the deviations 
are apreciable already at times which are about two orders of magnitude 
smaller than the ``intersection time''.
One should therefore be aware of this discrepancy when analyzing 
local quantities.

\begin{figure}
\epsfxsize=140mm
\hspace*{-10mmmm}
\epsffile{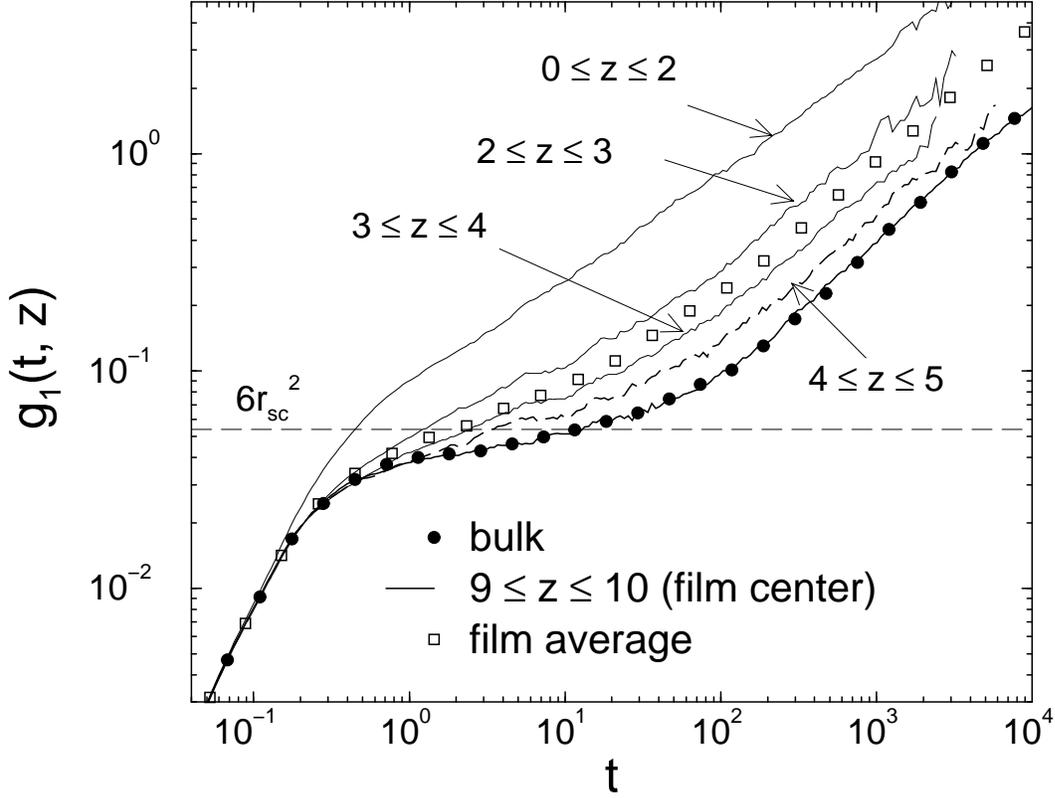}
\caption[]{Layer-resolved mean-square displacement of the innermost 
monomer, $g_1(z,t)$, versus time for 
a film of thickness $D\myeq 20$ ($\approx \! 14 R_\mr{g}$; $R_\mr{g}\myeq $ 
bulk radius of gyration) at $T\myeq 0.46$ 
($T_\mr{c}(D\myeq 20)\myeq 0.415$; see Table~\ref{tab1}). $z$ denotes the 
distance from the (left) wall. The 
displacements are calculated parallel to the wall and multiplied by $3/2$ 
(to put them on the same scale as the bulk data ($\bullet$) which are 
averaged over 3 spatial dimensions instead of over only 
2 for the film). The $g_1(z,t)$-data (indicated by lines) are 
averages over all monomers between some
$z_\mr{min} \leq z \leq z_\mr{max}$, which remain in the specified 
$z$-interval for all times shown 
in the figure. Due to the loss of the statistical accuracy 
at late times, the data is some times cut off 
at times with large statistical noise.

In the middle of the film ($9 \leq z \leq 10$) $g_1(z,t)$
coincides with $g_1(t)$ of the bulk, whereas it is much faster at 
the wall ($0\leq z \leq 2$).
The average behavior of the film (i.e., the average over all layers) 
is shown by $\Box$. The dashed horizontal line indicates the plateau value
$6r_{\rm sc}^2$ ($\simeq 0.054$) of a MCT-analysis 
performed in the bulk
\cite{Bennemann-Baschnagel-Paul::EPJB10,%
Aichele-Baschnagel::EurPhysJE::I,%
Aichele-Baschnagel::EurPhysJE::II}.
}
\label{fig:g1_T0.46_D20_of_z_and_bulk}
\end{figure}

Figure~\ref{fig:g1_T0.46_D20_of_z_and_bulk} depicts
the local MSD of the innermost monomer, $g_1(t;z)$,
for a film of thickness $D\myeq 20$ at a low temperature of $T\myeq 0.46$.
The displacements are calculated parallel to the walls using 
Eq.~(\ref{eq:local::MSD::correct::def}).
We see from this figure that the mobility in film center is 
practically equal to that of the bulk, whereas monomers in 
the proximity of the wall are much mobile. There is a gradual 
transition from the two step relaxation characteristic of the 
bulk at this temperature to a smooth crossover from microscopic
to free diffusive motion as the layers are closer to the wall.
Whereas the bulk-like dynamics for the film center is 
plausible -the monomer density profile is flat there and 
equal to $\rho_{\rm bulk}$
(see Fig.~\ref{fig:density_profiles_p1_D20_T_sym}), 
the continuous speeding-up of the dynamics in spite of 
the pronounced oscillations of the monomer density is 
less intuitive. It could be related to the following points:
An important factor is certainly the wall potential. This potential
is softer [see Eq.~(\ref{eq:def:uw})] than the LJ-potential which means 
that a monomer close to wall can further penetrate the wall than
a monomer can approach the center of mass of its nearest neighbors.
This should yield a higher mobility of the monomers at the wall 
at low $T$ where the dense local packing of the monomers is 
responsible for the mutual blocking and slowing down of the 
bulk dynamics.

On the other hand, the static structure factor of 
Fig.~\ref{fig:model} shows that the first maximum is not as 
large in the film as in the bulk at the same low temperature.
This implies that the overall arrangement of the monomers in the 
nearest neighbor shells is not as pronounced  in direction 
parallel to the wall as it is in the bulk. Since this local
packing is an important factor for the slowing down of the 
structural relaxation (Hansen-Verlet freezing criterion), 
the dynamics in the film should be faster.

Using a local version of Eq.~(\ref{eq:tau_gi::def}), we define a 
$z$-dependent relaxation time, $\tau(z)$,
\begin{eqnarray}
g_{i} (t \myeq \tau(z); z) &:=& 1 \mbox{~(defining equation for $\tau(z)$),}
\label{eq:tau_of_z_gi::def}
\end{eqnarray}
where $i\myeq 0, 1, 3,4$ denotes various types of displacements of
a polymer system [see 
sections~\ref{section::influence::of::ensembles::on::the::dynamics}
and~\ref{section:Dependence_of_Tc_on_Film_Thickness}].

\begin{figure}
\epsfxsize=140mm
\hspace*{-13mm}
\hspace*{15mm}\epsffile{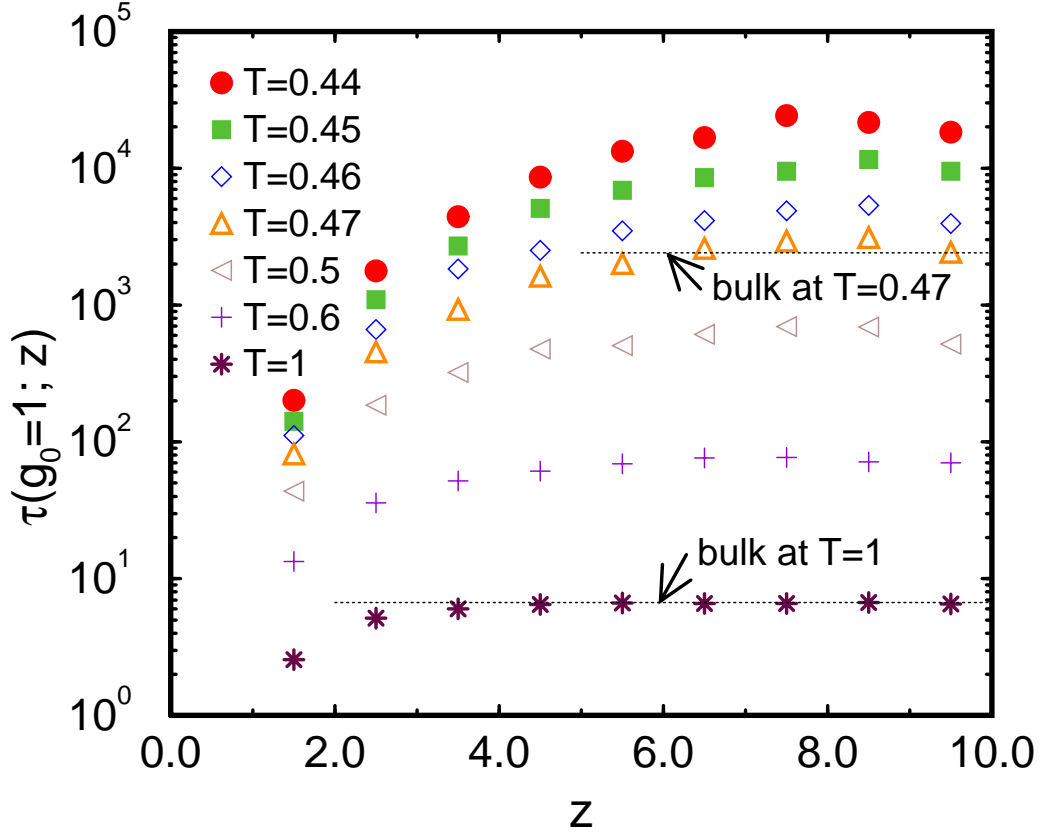}
\caption[]{Relaxation time $\tau(g_0 \myeq 1; z)$,
versus distance from the wall for a film of
thickness $D\myeq 20$ ($\approx \! 14 R_\mr{g}$; $R_\mr{g}\myeq $ bulk 
radius of gyration).
Here, $z$ stands for the distance  of the center of a layer (bin)
from the (left) wall. The thickness of a bin is $\Delta z \myeq 1$.
In the calculation of the local mean-square displacements of all 
monomers, $g_0(t, z)$,  which underlies the definition of $\tau$,
we consider the contribution of only those monomers, which remain 
within the specified layer for all times $t' \! \le \! t $
(see also the text). At high temperatures, there is a wide region 
around the film center, where $\tau$ is independent of $z$.
Contrary to that, at low $T$ the presence of the walls is ``felt''
also close to the film center 
[note that the lowest temperature is close to $T_\mr{c}(D\myeq 20)\myeq 0.415$; 
see Table~\ref{tab1}].
}
\label{tau_of_z_g0_eq0.667.D20}
\end{figure}

\begin{figure}
\epsfxsize=140mm
\hspace*{-3mm}
\epsffile{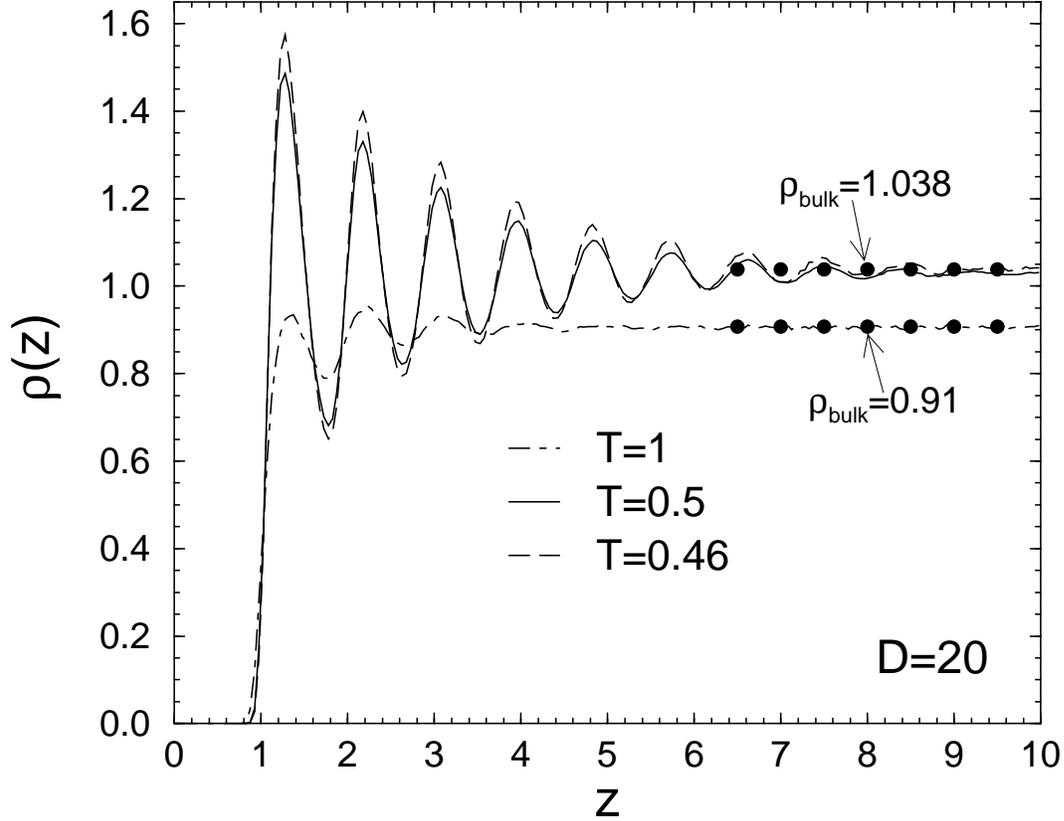}
\caption[]{Temperature dependence of the monomer density profile $\rho(z)$ 
for $D\myeq 20 \approx \! 14 R_\mr{g}$
($R_\mr{g}\myeq $ bulk radius of gyration). $z$ denotes the position of a 
monomer from the (left) wall. Since
the profiles are symmetric with respect to the middle of the film, only 
one half is shown. The temperatures
are characterisitc of the high-temperature state ($T\myeq 1$) and the 
supercooled state ($T\myeq 0.46 > T_\mr{c}(20) \myeq 
0.415$; see Table~\ref{tab1}) of the melt. The horizontal filled 
circles ($\bullet$) indicate the bulk
densities $\rho_\mr{bulk}\myeq 0.91$ and $\rho_\mr{bulk}\myeq 1.038$
at $T\myeq 1$ and $T\myeq 0.46$, respectively.}
\label{fig:density_profiles_p1_D20_T_sym}
\end{figure}

For a film of thickness  $D\myeq 20$, we applied
Eq.~(\ref{eq:tau_of_z_gi::def}) to the local MSD of all monomers, $g_{0}(t;z)$,
and computed $\tau(z)$ at various temperatures.
The results thus obtained are displayed in Fig~\ref{tau_of_z_g0_eq0.667.D20}.
Not unexpectedly, the effect of the walls is quite small at 
high temperatures. At $T\myeq 1$, for
example, there is a wide region of $z$-independent 
relaxation time around the film center. This constant value agrees with 
the relaxation time $\tau$ obtained by applying 
Eq.~(\ref{eq:tau_gi::def}) to the  bulk data.
As the temperature decreases the presence of the walls is ``felt'' 
also in the inner part of the film and the width of the 
region of constant relaxation time decreases.

This propagation of the wall effects into the inner part of the
film is also observed in the density profile.
Figure~\ref{fig:density_profiles_p1_D20_T_sym} shows that
density oscillations become more pronounced and 
long ranged at lower temperatures. Since the normal pressure
is kept constant during the simulations (as it was also the case in
preceeding simulations in the bulk~\cite{Bennemann-Baschnagel-Paul::EPJB10}),
the average value of the density in the film 
center increases with decreasing temperature and 
coincides with that of the bulk at the same temperature.
This increase of the density in the film gives raise to more 
pronounced oscillations of the monomer profile.

%%%%%%%%%%%%%%%%%%%%%%%%%%%%%%%%%%%%%%%%%%%%%%%%%%%%%%%%%%%%%%%%%%%%
%%%%%%%%%%%
%%%%%%%%%%%
\section{Conclusion}
\label{section:Conclusion}
%%%%%%%%%%%
%%%%%%%%%%%
%%%%%%%%%%%%%%%%%%%%%%%%%%%%%%%%%%%%%%%%%%%%%%%%%%%%%%%%%%%%%%%%%%%%
We presented results on molecular dynamics simulations for a model of 
non-entangled short polymer chains confined between two perfectly flat,
non-adsorbing and impenetrable walls. The monomer-wall interaction is 
modeled by potential which diverges as $z^{-9}$ when a monomer 
approaches the wall. This repulsion is weaker than the 12-6 Lennard-Jones 
potential of the monomer-monomer interaction and also than the soft-sphere
potential ($\sim z^{-12}$) used in simulations of confined supercooled 
simple liquids \cite{Yamamoto::JPhysIV::VOL::10::2000,FehrLoewen::PRE::1995}.
This special choice of the monomer-wall interaction has an influence on
both static and dynamic properties of the polymer films.

We find oscillations of the monomer density which start from a large value close
to the wall and decay towards the bulk density in the middle of the film for 
sufficiently large film thickness.  The amplitude of these oscillations becomes 
more pronounced with decreasing temperature, since the bulk density increases in 
our constant-pressure simulations.  However, the height of the largest maximum
at the wall does not exceed 1.5 times the bulk density in the temperature range
studied. This is much smaller than the difference found in simulations where
the particle-wall interaction is modeled by a soft-sphere potential. There, the
density can be more than 4 times larger than the bulk density at low $T$, which
leads to a vanishing of the density in the subsequent minimum. In this case it is 
conceivable that the smoothness of the walls, which could promote fast particle 
motion (``slip boundary condition''), is completely outweighed by the high
density in the first layer, and one can speculate that the subsequent layers are also slowed
down with respect to the bulk. This could explain why the studies 
\cite{Yamamoto::JPhysIV::VOL::10::2000,FehrLoewen::PRE::1995} found an increase
of the glass transition temperature rather than a decrease although the confinement
was realized by completely smooth walls as in our work.

In our work the repulsive monomer-wall interaction is much softer, even than that
between the monomers in the bulk.  Therefore, a monomer can come closer to the wall
than to anyone of its nearest-neighbors.  This should give a monomer at the wall  
more freedom to move. Since the slowing down of the structural relaxation of our
model in the bulk mainly results from the blocking of a particle by its nearest
neighbors \cite{Bennemann-Baschnagel-Paul::EPJB10,Aichele-Baschnagel::EurPhysJE::I::D,Aichele-Baschnagel::EurPhysJE::II::D}, one can expect this difference in repulsive interaction
to become particularly important at low temperatures close to $T_{\rm c}$. 
This expectation is borne out by the simulation data. Whereas the monomers in 
the film center exhibit very sluggish motion in an intermediate time interval,
a signature of the ``cage effect'', this intermittence of the displacements is 
not at all visible for the monomers in contact with the wall. They behave as if
they were at a higher temperature (see Fig.~\ref{fig:g1_T0.46_D20_of_z_and_bulk}).  
Their higher mobility also triggers faster motion of adjacent monomers, leading
to an overall acceleration of the dynamics in the film.

We have tried to quantify this acceleration by fitting various relaxation times
and the diffusion coefficient of a chain to both a power law, motivated by
mode-coupling theory (MCT), and by the Vogel-Fulcher-Tammann (VFT) equation.  The 
power law yields an estimate of the critical temperature, $\Tc$, and the VFT-equation
of the Vogel-Fulcher temperature, $T_0$, as a function of film thickness $D$.  In
accord with the qualitative observations described above both $\Tc(D)$ and $T0(D)$
decrease with decreasing film thickness. Furthermore, a comparison of the rescaled 
quantities, i.e., of $\Tc(D)/ \Tcbulk$ and $\Tnull(D)/ \Tnullbulk$, shows that the 
variation with $D$ of $\Tc(D)$ agrees with that of $\Tnull(D)$.  Although the
extrapolation to $\Tnull$ is less reliable than that to $\Tc$, since it has to 
cover a larger $T$-range, this result suggests that the glass transition 
temperature of our model should follow the same behavior.

The reduction of the critical temperature seems to be the dominant influence of
the walls on the dynamics in the time window of the MCT-$\beta$ process. For 
these intermediate times, the mean-square displacements of the films of various 
thicknesses and of the bulk exhibit the same time dependence when compared for a 
fixed reduced temperature $T \! - \! \Tc$.  This shifting property of the temperature 
axis suggests that similar processes cause the intermittence of monomer motion on 
these intermediate time scales in the bulk and in the films. However, deviations 
between film and bulk dynamics are found as soon as the monomers escape from 
their local environment and the $\alpha$-relaxation sets in.  In order to 
understand these differences better we want to investigate the incoherent 
intermediate scattering function and related quantities which provide 
information on the dynamics on different length scales. Work in this 
direction is under way.

%%%%%%%%%%%%%%%%%%%%%%%%%%%%%%%%%%%%%%%%%%%%%%%%%%%%%%%%%%%%%%%%
% ------------------------------------------------
% Acknowledgement
% ------------------------------------------------
\section*{Acknowledgement}
We thank J. Horbach, P. Scheidler, C. Brangian and M. Aichele for helpful 
discussions on various aspects of this work. We gratefully 
acknowledge the financial support by the 
``Deutsche Forschungsgemeinschaft'' (DFG) under the project number 
SFB262 and by BMBF under the project number 03N6015.
We are also indebted to the European Science 
Foundation for financial support by the ESF Programme on ``Experimental and 
Theoretical Investigations of Complex Polymer Structures'' (SUPERNET).
Generous grants of simulation time by 
the computer center at the university of Mainz (ZDV), 
the NIC in J\"ulich and 
the RHRK in Kaiserslautern are also acknowledged.

% ------------------------------------------------
% References
% ------------------------------------------------
% ------------------------------------------------
% Tables
% ------------------------------------------------
\newpage
\begin{table}
\begin{tabular}{|c||c|c|c|c|c|c|}
$D$        & 5     & 7     & 10    & 15    & 20    & bulk \\
\hline
$\Tnull$ & 0.204 $\pm$ 0.007 & 0.253 $\pm$ 0.013 & 0.288 $\pm$ 0.006 & 0.297 $\pm$ 0.007 
& 0.308 $\pm$ 0.004 & 0.328$\pm$ 0.008 \\
\hline
$T_\mr{c}$ & 0.305 $\pm$ 0.006 & 0.365 $\pm$ 0.007 & 0.390 $\pm$ 0.005 & 0.405 $\pm$ 0.008 
& 0.415 $\pm$ 0.005 & 0.450 $\pm$ 0.005 \\
\hline
$\gamma$ & 2.5 $\pm$ 0.2 & 2.4 $\pm$ 0.2 & 2.1 $\pm$ 0.1 & 2.2 
$\pm$ 0.1 & 2.1 $\pm$ 0.1 & 2.09
\end{tabular}
\caption[]{Survey of the VFT-temperature, $\Tnull$, mode-coupling 
critical temperature, $\Tc$, and the 
critical exponent, $\gamma$, for different film thicknesses $D$ and for the bulk.
$\Tnull$ was determined via fits to Eq.~(\ref{eq:VFT::law::for::tau}) both for the 
film and for the bulk. As to $\Tc$, we determined $\Tc(D)$ from fits to 
Eq.~(\ref{eq:MCT::power::law::for::tau}). $T^{\rm bulk}_{\rm c}$ was known from 
previous analyses~\cite{Bennemann-Baschnagel-Paul::EPJB10,%
Aichele-Baschnagel::EurPhysJE::I,%
Aichele-Baschnagel::EurPhysJE::II}. 
This  result for  $T^{\rm bulk}_{\rm c}$ is also obtained 
applying Eq.~(\ref{eq:MCT::power::law::for::tau}) to the bulk data.
}
\label{tab1}
\end{table}
%
% ------------------------------------------------
% Figures and figure captions
% ------------------------------------------------
\vspace*{-20mm}
%%
%%
%%
%%%%%%%%%%%%%%%%%%%%%%%%%%%%%
%%%%%%%%%%%%%%%%%%%%%%%%%%%%%%
%%
%%
%%
\newpage
%%
%%
%%%%%%%%%%%%%%%%%%%%%%%%%%%%%%
%%%%%%%%%%%%%%%%%%%%%%%%%%%%%%%%%%%%
%%
%%
\newpage
%
%
%%%%%%%%%%%%%%%%%%%%%%%%%%%%%%%%%%%%%
%%%%%
%%
\newpage
%%
%%
%%%%%%%%%%%%%%%%%%%%%%%%%%%%%%%%%%%
\newpage
\newpage
%%
%%%
\newpage
%%
%%%
\newpage

\newpage
\newpage
%%
%%%%%%%%%%%%%%%%%%%%%%%%%%%%%%%%%%%%%%%%%%
\newpage
\newpage
\newpage
\newpage
\vspace*{-20mm}
%%
%%
%%%%%%%%%%%%%%%%%%%%%%%%%%%%%%%%%%%%%%%%%%%%%%%%%%%%%%%%%%
%%
%%
\newpage
%%%%%%%%%%%%%%%%%%%%%%%%%%%%%%%%%%%%%%%%%%%%%%%%%%%%%%%%%%
%%
%%
\newpage
%%%%%%%%%%%%%%%%%%%%%%%%%%%%%%%%%%%%%%%%%%%%%%%%%%%%%%%%%%%%%
%%
%%
%%
\newpage
%%
%%
%%%%%%%%%%%%%%%%%%%%%%%%%%%%%%%%%%%%%%%
%%
%%
%%%%%%%%%%%%%%%%%%%%%%%%%%%%%%%%%%%%%%%
%%
\newpage

%================================================================
\end{document}